\documentclass[a4paper,11pt,english]{article}

\usepackage{amsfonts,amssymb,amsthm}    
\usepackage[affil-it,auth-sc]{authblk}          
            
\usepackage{bigints}
\usepackage{bm}                         
\usepackage{braket}              
\usepackage[utf8]{inputenc}
\usepackage{caption}
\usepackage{enumerate}
\usepackage{enumitem}           
\usepackage[left=22mm,right=22mm,top=22mm,bottom=24mm]{geometry}           
\usepackage{latexsym}            
\usepackage{mathrsfs}           
\usepackage{mathtools}                  
\usepackage{mparhack,fixltx2e,relsize}     
\usepackage[only,llbracket,rrbracket]{stmaryrd} 
\usepackage{subfig}                      
\usepackage{cite}
\usepackage{upgreek}
\usepackage[greek,english]{babel}                   
\usepackage{graphicx}
\usepackage{color}
\usepackage[titletoc,toc,title]{appendix}
\usepackage{lmodern}
\usepackage{multirow}
\usepackage{float}

\usepackage[english]{varioref}
\usepackage[hidelinks]{hyperref}
\usepackage{bookmark}

\hypersetup{
  breaklinks=true,
  bookmarksnumbered,
  bookmarksopen=true, 
  bookmarksopenlevel=2, 
  pdftitle={Scattering, Trapping and Cloaking-Type Effects of Plane Waves by Point Scatterers in Strain Gradient Elasticity},
  pdfauthor={E. Alevras, Th. Zisis, Panos A. Gourgiotis},
  pdfkeywords={dispersion; Green's function; metamaterials; microstructure; resonance} }

\begin{document}
\selectlanguage{english}

\title{Scattering, Trapping and Cloaking-Type Effects of Plane Waves by Point Scatterers in Strain Gradient Elasticity}

\author[1]{E. Alevras}
\author[1]{Th. Zisis}
\author[1]{P.A. Gourgiotis\footnote{Corresponding author:\,e-mail:\,pgourgiotis@mail.ntua.gr.}}
\affil[1]{Department of Mechanics, School of Applied Mathematical and Physical Sciences, National Technical University of Athens, Zographou, 15773, Greece}

\date{}
\maketitle

\begin{abstract}
\noindent
Wave scattering by localized constraints in microstructured solids is strongly affected by the interplay between material length scales, dispersion and geometry. This work investigates plane-strain scattering of time-harmonic P and SV waves by clusters of rigid point constraints embedded in an infinite medium governed by strain gradient elasticity. A closed-form dynamic Green's tensor is derived for the plane-strain problem. In contrast to the classical elastodynamic Green's tensor, the strain gradient Green's tensor remains bounded at the source, allowing point constraints to be introduced directly through a superposition of fundamental solutions. The resulting multiple-scattering problem is reduced to a finite-dimensional algebraic system for the reaction amplitudes exerted by the pins. A frequency-domain procedure is developed to identify resonance-like amplification and trapping. Candidate resonant frequencies are first associated with local minima of the determinant of the Green's matrix, while higher-order curvature criteria are used to distinguish trapping-dominated resonances from ordinary non-localized scattering responses. The results show that the response is governed primarily by the microstructural ratio between the microinertial and energetic strain gradient lengths. In the anomalous dispersion regime, sharp resonant minima lead to strong displacement localization within the pin configuration, including perimeter-localized trapping modes in dense circular arrays. In the normal dispersion regime, these resonances are strongly attenuated and the pins act as weak scatterers, giving rise to a cloaking-type response in which the incident field is only weakly perturbed. The influence of Poisson’s ratio, incidence angle and compound pin configurations is also examined. The analysis reveals how intrinsic material lengths and geometric arrangement can be used jointly to tune scattering, trapping and wave-screening mechanisms in microstructured elastic media.
\end{abstract}


\section{Introduction}
\noindent
Wave propagation in microstructured solids is governed not only by elastic moduli and mass density, but also by internal length scales associated with the underlying architecture. These length scales become important when the wavelength is comparable to microstructural features, giving rise to dispersion, boundary-layer effects, localization and other scale-dependent phenomena that cannot be captured by classical elastodynamics. Such effects are central to the mechanics of architected materials, phononic crystals and elastic metamaterials, where wave propagation can be shaped by geometry and microstructure rather than by material contrast alone \cite{craster2012acoustic, kadic2013metamaterials, romeo2013wave, madeo2015generalized, rosi2016anisotropic}. Generalized continuum theories, and in particular strain gradient elasticity, provide a continuum framework in which these effects can be incorporated naturally through higher-order kinematics and intrinsic material lengths.

The present work addresses a fundamental scattering problem in this setting: the interaction of plane P and SV waves with clusters of rigid point constraints embedded in an infinite strain gradient elastic medium under plane-strain. The point constraints may be viewed as idealized models of very stiff, localized microstructural inclusions or anchoring sites in a softer dispersive elastic matrix. Physical analogues include microfabricated elastic metamaterials containing rigid fibres, biological materials such as cortical bone where osteonal structures introduce intrinsic length scales and local stiffness contrasts, and soft tissues or hydrogels containing stiff inclusions, cell nuclei or microcalcifications \cite{aggelis2004iterative, maurel2006propagation, vavva2009velocity, papacharalampopoulos2011numerical, charalambopoulos2012gradient, morini2014remarks, muhlestein2018acoustic, nobili2019diffraction, nobili2020new}. In all these systems, elastic or ultrasonic waves interact with spatially distributed microstructural obstacles, producing multiple scattering, local amplification and, under suitable conditions, wave localization or weak-scattering responses. Circular or clustered arrangements of such inclusions therefore provide a canonical setting for studying how material dispersion and geometry combine to control wave motion, with potential relevance to engineered wave manipulation in metamaterials and microstructured media \cite{brule2014experiments, colombi2016seismic, colombi2016forests, colombi2017elastic, de2020graded}.

The present formulation builds on point-scatterer methods originally developed by Evans \& Porter \cite{evans2007penetration} and on related approaches for pinned Kirchhoff plates \cite{reddy2006theory, smith2012negative, haslinger2012transmission, haslinger2018localization, lazaro2026weak}. These scalar scattering frameworks have proved effective for modelling wave interaction with arrays of point constraints, and were subsequently adapted by Alevras \textit{et al.} \cite{alevras2026scattering} to strain gradient elasticity under anti-plane shear conditions. That adaptation relied partly on formal analogies between classical plate theory and simplified gradient-elastic models \cite{gavardinas2018karman}. However, such analogies are limited in scope and do not capture the full vectorial character of plane-strain motion, where P and SV waves coexist, mode conversion occurs and the Green's function has a tensorial form. The present work therefore extends point-scatterer ideas to a genuinely vectorial strain gradient elastodynamic setting.

The use of point constraints in classical two-dimensional elastodynamics is limited by the singular character of the classical Green's tensor at the source. In strain gradient elasticity, however, the presence of the energetic length scale regularizes the displacement Green's tensor, which remains bounded at the point of application of a concentrated force. This property makes it possible to formulate a point-scatterer problem directly, without assigning a finite radius to the inclusions. The resulting model should not be regarded as the small-radius limit of classical elasticity in a trivial sense. Rather, it exploits a genuinely different regularized point-constraint limit enabled by the gradient theory. A comparison with classical elastodynamics containing small but finite rigid inclusions would therefore be a natural extension, and would help separate localization mechanisms due to multiple scattering alone from those governed by intrinsic material length scales.

The mathematical foundation of the present study is the derivation of a closed-form, time-harmonic Green's tensor for strain gradient elasticity under plane-strain. The Green's tensor is obtained by a double Fourier transform and contains both outgoing cylindrical waves and exponentially decaying evanescent contributions. The propagating part is associated with the real roots of the P and SV-wave dispersion relations, whereas the evanescent part is associated with purely imaginary roots. The boundedness of the displacement Green's tensor at the source allows the total wave field generated by an arbitrary cluster of pins to be represented as the superposition of the prescribed incident field and the fields radiated by fictitious concentrated forces placed at the pin locations.

This construction leads to a finite-dimensional, non-homogeneous algebraic system for the unknown reaction amplitudes exerted by the pins. The Green's matrix of this system depends on the frequency, the material parameters and the pin geometry, while the incident wave enters through its values sampled at the pin locations. Resonant behavior is therefore interpreted in a generalized forced-response sense: it corresponds to frequencies at which the prescribed incident field, through its interaction with the pin configuration, produces strong scattered-field amplification. Candidate resonant frequencies are identified from local minima of the determinant of the Green's matrix. Since not every such minimum produces a trapped field, higher-order curvature criteria are introduced to distinguish sharp minima associated with trapping-dominated resonances from ordinary non-localized scattering responses.

A central result of the present work is that the qualitative nature of the response is governed primarily by the dimensionless microstructural ratio $H=h/\ell$, where $h$ is the microinertial length and $\ell$ is the energetic strain gradient length. These internal lengths play a fundamental role in shaping the system's dynamic response, governing both the dispersive nature of wave propagation and the interaction of waves with localized constraints \cite{madeo2015wave, rosi2016anisotropic}. When $H<1$, the medium exhibits anomalous dispersion and the system supports sharp resonant minima leading to strong localization of the displacement field within the pinned region. For dense circular arrays, the localization may become concentrated along the pinned perimeter, producing a boundary-localized or whispering-gallery-type trapping response. When $H>1$, corresponding to normal dispersion, these trapping resonances are strongly weakened and the pins behave as weak scatterers over the frequency range examined. In this regime, the response has a cloaking-type character: the incident wave field is only weakly perturbed by the pin configuration. This effect does not arise from coordinate transformations \cite{greenleaf2003nonuniqueness, greenleaf2003anisotropic, pendry2006controlling, leonhardt2006optical}, as in transformation elastodynamics \cite{milton2006cloaking, colquitt2014transformation, brun2014transformation}, but from the interplay between strain gradient dispersion, microinertia and pin geometry. The case $H=1$ forms a transitional non-dispersive limit in which the propagating waves travel at the classical phase velocities, while the Green's tensor retains the regularizing evanescent contribution of the strain gradient theory.

The analysis also examines the influence of Poisson's ratio, incident-wave polarization, incidence angle and compound pin geometries. Poisson's ratio affects the Green's matrix through the P/SV wave-speed contrast and modifies the coupling between the incident field and localized response patterns. The angle of incidence does not alter the determinant-based candidate resonant frequencies, but changes the sampled incident-field vector and therefore the amplitude with which a given resonant pattern is excited. Multiple displaced or offset circular configurations introduce inter-cluster coupling and can produce collective localization, selective trapping or enhanced confinement through repeated scattering. These effects highlight the potential of generalized continuum models to capture unconventional dynamic responses in microstructured solids \cite{gourgiotis2017dynamics, bigoni2016folding}. Beyond the specific configurations studied here, the analytical framework can be extended to other generalized continuum theories, including couple-stress elasticity and micromorphic models \cite{toupin1964theories, eringen1968mechanics, eringen2012microcontinuum, neff2014unifying}, and may provide a basis for designing microstructured media with tailored capabilities for vibration control, wave redirection and energy harvesting enhancement \cite{delfani2013enhanced, brule2014experiments, colombi2016seismic, delfani2017elastic, colombi2016forests, colombi2017elastic, de2020graded}. Furthermore, these generalized continuum approaches are increasingly utilized to characterize dynamic anti-plane slip pulses, super-shear fracture, and Mach-type wave patterns in microstructured flexoelectric media \cite{majdoub2008enhanced, ahmadpoor2015flexoelectricity, giannakopoulos2023hyperbolicityI, giannakopoulos2023hyperbolicityII, knisovitis2024anti, giannakopoulos2025failure}.

The paper is organized as follows. In \hyperref[sec2]{Section~2}, we summarize the governing equations of strain gradient elastodynamics, with emphasis on the dispersion relations and the role of the two characteristic lengths. In \hyperref[sec3]{Section~3}, we derive the time-harmonic plane-strain Green's tensor and discuss its regularity at the source. In \hyperref[sec4]{Section~4}, we formulate the point-scatterer problem for arbitrary pin clusters and introduce the frequency-domain criteria used to identify trapping resonances. In \hyperref[sec5]{Section~5}, we present the parametric analysis of circular, displaced and offset pin configurations, highlighting the transition between trapping and cloaking-type responses. Concluding remarks are given in \hyperref[sec6]{Section~6}.

\section{Fundamentals of strain gradient elastodynamics}
\label{sec2}
\noindent
In this section, we briefly present the basic elastodynamic equations of the Toupin-Mindlin strain gradient elasticity \cite{toupin1962elastic, mindlin1964micro}. If a continuum with microstructure is viewed as a collection of subparticles (micro-media) having the form of unit cells (cubes), the following expression of the kinetic energy density $\mathcal{T}$ is obtained with respect to a Cartesian coordinate system $Ox_{1}x_{2}x_{3}$ (Mindlin \cite{mindlin1964micro}):
\begin{equation}\label{eq:2.1}
\mathcal{T} = \frac{1}{2} \rho \dot{u}_{i} \dot{u}_{i} + \frac{1}{2} \rho h^2 \left( \partial_{i} \dot{u}_{j} \right) \left( \partial_{i} \dot{u}_{j} \right),
\end{equation}
where $\rho$ is the density per unit volume, and $h$ is a microstructural length associated with the microinertia of the medium, $u_{j}$ is the displacement vector, $\partial_{i}(\cdot) \equiv \partial(\cdot)/\partial x_{i}$, the superposed dot denotes time derivative, and the Latin indices span the range $(1, 2, 3)$ (an indicial notation and summation convention is used throughout). The second term in the right-hand side of Eq.~\eqref{eq:2.1}, involving the velocity gradients, represents the microinertia of the continuum. This term, which is not encountered within classical continuum mechanics, reflects the more detailed description of motion in the present theory. According to Mindlin \cite{mindlin1964micro}, $h^2 = d^2 / 3$, where $d$ is the side length of the representative cell of the microstructure.

Next, the following expression of the strain energy density is postulated for a centrosymmetric material:
\begin{equation}\label{eq:2.2}
\mathcal{W} = \frac{1}{2} c_{ijkl} \varepsilon_{ij} \varepsilon_{kl} + \frac{1}{2} d_{ijklmn} \kappa_{ijk} \kappa_{lmn},
\end{equation}
where $(c_{ijkl}, d_{ijklmn})$ are tensors of material constants, $\varepsilon_{ij} = (1/2) (\partial_{i} u_{j} + \partial_{j} u_{i}) = \varepsilon_{ji}$ is the linear strain tensor and $\kappa_{ijk} = \partial_{i} \varepsilon_{jk}$ is the strain gradient. The form in \eqref{eq:2.2} can be viewed as a more accurate description of the constitutive response than that provided by the classical elasticity.

Then, appropriate definitions for the stresses follow as
\begin{equation}\label{eq:2.3}
\tau_{ij} \equiv \frac{\partial \mathcal{W}}{\partial \varepsilon_{ij}}, \qquad \mu_{ijk} \equiv \frac{\partial \mathcal{W}}{\partial \kappa_{ijk}} = \frac{\partial \mathcal{W}}{\partial_{i} \varepsilon_{jk}},
\end{equation}
where $\tau_{ij}$ is the monopolar stress tensor, and $\mu_{ijk}$ is the dipolar (or double) stress tensor (a third-rank tensor) expressed in dimensions of [force][length]$^{-1}$. The dipolar stress tensor follows from the notion of dipolar forces, which are anti-parallel forces acting between the micro-media contained in the continuum with microstructure \cite{jaunzemis1967continuum}. According to \eqref{eq:2.3}, the following symmetries for the monopolar and dipolar stress tensors are noticed: $\tau_{ij} = \tau_{ji}$ and $\mu_{ijk} = \mu_{ikj}$.

The equations of motion and the boundary conditions can be obtained from Hamilton's principle and variational considerations using \eqref{eq:2.1} and \eqref{eq:2.2}. The variational form of Hamilton's principle assumes the following form (see \cite{mindlin1964micro} for a slightly different formulation)
\begin{align}\label{eq:2.4}
\int_{t_{1}}^{t_{2}} \left[ \int_{V} \left( \delta \mathcal{T} - \delta \mathcal{W} \right) dV \right] dt
&+ \int_{t_{1}}^{t_{2}} \int_V F_{q} \delta u_{q} dV dt
+ \int_{t_{1}}^{t_{2}} \int_{V} \Phi_{qr} \delta \partial_{r} u_{q} dV dt \nonumber \\
&+ \int_{t_{1}}^{t_{2}} \left[ \int_{S} t^{(n)}_{q} \delta u_{q} dS 
+ \int_S T^{(n)}_{qr} \partial_{q} \left( \delta u_{r} \right) dS \right] dt = 0,
\end{align}
where $V$ is the region occupied by the body and $\partial V$ denotes any boundary along a section inside the body or along the surface of it. The symbol $\delta$ denotes weak variations, and it acts on the quantity existing on its right. Also, $t_{1}$ and $t_{2}$ are two arbitrary instants of time for which the variations $\delta u_{i}$ are zero at all points of the body. In addition, $F_{q}$ is the monopolar body force, $\Phi_{qr}$ is the dipolar (or double) body force, $t^{(n)}_{q}$ is the true monopolar traction, $T^{(n)}_{pq}$ is the true dipolar traction, and $n_{p}$ is the outward unit vector, normal to the boundary.

In the present study, dipolar body forces are neglected; however, they are retained in the formulation of Hamilton's principle for completeness. It then follows that the equations of motion and the corresponding traction boundary conditions on a smooth boundary are given by \cite{Eshel}:
\begin{equation}\label{eq:2.5}
\partial_{j} \left( \tau_{jk} - \partial_{i} \mu_{ijk} \right) + F_{k} = \rho \ddot{u}_{k} - \rho h^2 \partial_{jj} \ddot{u}_{k} \quad \text{in } V,
\end{equation}
\begin{equation}\label{eq:2.6}
P_{k}^{(n)} = n_{j} (\tau_{jk} - \partial_{i} \mu_{ijk}) - D_{j} \left( n_{i} \mu_{ijk} \right) + \left( D_{p} n_{p} \right) n_{i} n_{j} \mu_{ijk} + \rho h^2 n_{j} \partial_{j} \ddot{u}_{k} \quad \text{on } \partial V,
\end{equation}
\begin{equation}
\label{eq:2.7}
R_{k}^{(n)} = n_{i} n_{j} \mu_{ijk} \quad \text{on } \partial V,
\end{equation}
where $D_{i}(\cdot) \equiv \partial_{i}(\cdot) - n_{i} D(\cdot)$ is the surface gradient operator, and $D(\cdot) \equiv n_{i} \partial_{i}(\cdot)$ is the normal gradient operator. The auxiliary force traction $P^{(n)}_{i}$ and the auxiliary double force traction $R^{(n)}_{i}$ are related to the true force traction $t^{(n)}_{i}$ and the true double force traction $T^{(n)}_{ij}$ through the relations: $P_i^{(n)} \equiv t_i^{(n)} + (D_k n_k)\, n_j T_{ji}^{(n)} - D_j T_{ji}^{(n)}$ and $R_i^{(n)} \equiv n_j T_{ji}^{(n)}$ (see \cite{bleustein1967note}). Also, the pertinent kinematical boundary conditions of the theory were derived by Mindlin \cite{mindlin1964micro} (see also \cite{georgiadis2006energy}), but are omitted here since they are not relevant to our specific problem.

In our study, we consider the simplest possible linear and isotropic form of the strain energy density function (\cite{georgiadis2003high, lazar2005nonsingular}), i.e.,
\begin{equation}\label{eq:2.8}
\mathcal{W} = \frac{1}{2} \lambda \varepsilon_{ii} \varepsilon_{jj} + \mu \varepsilon_{ij} \varepsilon_{ij} + \frac{1}{2} \lambda \ell^2 \left( \partial_{k} \varepsilon_{ii} \right) \left( \partial_{k} \varepsilon_{jj} \right) + \mu \ell^2 \left( \partial_{k} \varepsilon_{ij} \right) \left( \partial_{k} \varepsilon_{ij} \right),
\end{equation}
where $\ell^2$ is the gradient coefficient having dimensions of [length]$^2$, and $(\lambda, \mu)$ are the standard Lam\'e constants with dimensions of [force][length]$^{-2}$. In this way, only one new material constant is introduced in the strain energy density. Notice that the general expression for the strain energy density in the isotropic case involves five additional material constants besides the two Lam\'e constants \cite{mindlin1964micro}. Combining now Eqs.~\eqref{eq:2.3} with \eqref{eq:2.8} provides the following constitutive equations:
\begin{equation}\label{eq:2.9}
\tau_{ij} = \lambda \delta_{ij} \varepsilon_{kk} + 2 \mu \varepsilon_{ij}, \qquad \mu_{ijk} = \ell^2 \partial_{i} \left( \lambda \delta_{jk} \varepsilon_{qq} + 2 \mu \varepsilon_{jk} \right) = \ell^2 \partial_{i} \tau_{jk},
\end{equation}
where $\delta_{ij}$ is the Kronecker delta. The particular choice of \eqref{eq:2.8} is physically justified and possesses a notable symmetry. Indeed, Eqs.~\eqref{eq:2.9} indicate that this simple form of strain gradient elasticity is a strain gradient theory as well as a stress gradient theory. Finally, the restriction of positive definiteness of $\mathcal{W}$ requires the following inequalities for the material constants: $(3 \lambda + 2 \mu) > 0$, $\mu > 0$, $\ell^2 > 0$.

In summary, Eqs.~\eqref{eq:2.5}-\eqref{eq:2.7} and \eqref{eq:2.9} are the governing equations for the isotropic linear gradient elastodynamics. Combining \eqref{eq:2.5} with \eqref{eq:2.9}, one obtains the equations of motion in terms of displacements:
\begin{equation}\label{eq:2.10}
\left( 1 - \ell^2 \nabla^2 \right) \left[ \mu \nabla^2 \mathbf{u} + \left( \lambda + \mu \right) \nabla \left( \nabla \cdot \mathbf{u} \right) \right] + \mathbf{F} = \rho \left( \mathbf{\ddot{u}} - h^2 \nabla^2 \mathbf{\ddot{u}} \right),
\end{equation}
where $\nabla^{2}(\cdot)$ is the three-dimensional Laplace operator. In the limit $(\ell, h) \rightarrow 0$, the Navier-Cauchy equations of classical linear isotropic elasticity are recovered from \eqref{eq:2.10}. The fact that the gradient coefficient $\ell$ multiplies the higher-order term reveals the singular-perturbation character of the gradient theory and the emergence of associated boundary-layer effects. It is worth noting that Eq.~\eqref{eq:2.10} is of the same form as the linear version of the 'good' Boussinesq equation employed in the study of bidirectional solitary waves propagating on the free surface of a constant-depth fluid (see e.g., Maugin \cite{maugin1999nonlinear}).

Next, by taking the divergence and curl of the homogeneous Eq.~\eqref{eq:2.10} we obtain the equations governing the propagation of dilatation and rotation, respectively:
\begin{equation}\label{eq:2.11}
c^2_{P} \left( 1 - \ell^2 \nabla^2 \right) \nabla^2 \left( \nabla \cdot \mathbf{u} \right) = \left( 1 - h^2 \nabla^2 \right) \nabla \cdot \mathbf{\ddot{u}},
\end{equation}
\begin{equation}\label{eq:2.12}
c^2_{S} \left( 1 - \ell^2 \nabla^2 \right) \nabla^2 \left( \nabla \times \mathbf{u} \right) = \left( 1 - h^2 \nabla^2 \right) \nabla \times \mathbf{\ddot{u}},
\end{equation}
where $c_P = [( \lambda + 2 \mu )/ \rho]^{1/2}$ and $ c_{S} = \left( \mu/ \rho \right)^{1/2}$ are the propagation velocities of the pressure (P) and shear (S) waves, respectively, in the classical (i.e., non-gradient) elasticity theory. Moreover, we note that, unlike the corresponding case of classical elastodynamics, the partial differential equations (PDEs) \eqref{eq:2.11} and \eqref{eq:2.12} are of the fourth order. The fourth-order character of the field equations is associated with dispersive wave propagation, as shown below through the corresponding plane-wave dispersion relations.

The last statement can easily be supported by considering time-harmonic plane-wave solutions and determining dispersion relations. First, we consider a plane-wave solution of Eqs.~\eqref{eq:2.11} and \eqref{eq:2.12} in the following form:
\begin{equation}\label{eq:2.13}
\mathbf{u} = B \mathbf{d} e^{\left[ i (q \left(\mathbf{n} \cdot \mathbf{r}\right) - \omega t )\right]},
\end{equation}
where $B$ denotes the amplitude, $(\mathbf{d}, \mathbf{n})$ are unit vectors defining the directions of motion and propagation, respectively, $\mathbf{r}$ is the position vector, $q$ is the wavenumber, $\omega$ is the circular frequency of the plane wave, and $i^2=-1$. Then, on substituting \eqref{eq:2.13} into Eqs.~\eqref{eq:2.11} and \eqref{eq:2.12}, we obtain the following dispersion relations for the pressure and shear waves:
\begin{equation}\label{eq:2.14}
\omega^2 = c^2_{P} q^2 \left( 1 + \ell^2 q^2 \right) \left( 1 + h^2 q^2 \right)^{-1},
\end{equation}
\begin{equation}\label{eq:2.15}
\omega^2 = c^2_{S} q^2 \left( 1 + \ell^2 q^2 \right) \left( 1 + h^2 q^2 \right)^{-1}.
\end{equation}
Accordingly, the phase velocities of the longitudinal and shear waves in form II of Mindlin's general theory take the following forms:
\begin{equation}\label{eq:2.16}
V_{P} \equiv \frac{\omega}{q} = c_{P} \left( 1 + \ell^2 q^2 \right)^{1/2} \left( 1 + h^2 q^2 \right)^{-1/2},
\end{equation}
\begin{equation}\label{eq:2.17}
V_{S} \equiv \frac{\omega}{q} = c_{S} \left( 1 + \ell^2 q^2 \right)^{1/2} \left( 1 + h^2 q^2 \right)^{-1/2}.
\end{equation}
Eqs.~\eqref{eq:2.16} and \eqref{eq:2.17} show that the propagation velocities of these waves depend on the respective wavenumber. Hence, both waves are dispersive in form II of Mindlin's general theory. It is worth noting that a similar situation is encountered in coupled thermoelasticity with thermal relaxation, where exponentially decaying plane waves of pressure and shear type propagate with dispersion into the medium \cite{brock2011plane}. Note also that in couple stress elasticity only the shear waves are dispersive \cite{mindlin1962effects, gourgiotis2016stress}.

To investigate further the nature of the dispersion relations in form II of Mindlin's general theory, we consider the group velocity $V^g = d \omega / d q$ at which the energy propagates in a dispersive medium \cite{achenbach2012wave}. In particular, according to Eqs.~\eqref{eq:2.14}-\eqref{eq:2.17} we obtain
\begin{equation}\label{eq:2.18}
V^g_{P} = V_{P} + c_{P} \left( \ell^2 - h^2 \right) q^2 \left( 1 + \ell^2 q^2 \right)^{-1/2} \left( 1 + h^2 q^2 \right)^{-3/2},
\end{equation}
\begin{equation}\label{eq:2.19}
V^g_{S} = V_{S} + c_{S} \left( \ell^2 - h^2 \right) q^2 \left( 1 + \ell^2 q^2 \right)^{-1/2} \left( 1 + h^2 q^2 \right)^{-3/2}.
\end{equation}
The following three cases are then distinguished: $(i)$ For $\ell^2 < h^2$, Eqs.~\eqref{eq:2.18} and \eqref{eq:2.19} imply that $V^g_{P} < V_{P}$ and $V^g_{S} < V_{S}$, thus the dispersion is normal. $(ii)$ For $\ell^2 > h^2$, we have $V^g_{P}>V_{P}$ and $V^g_{S}>V_{S}$ indicating that the dispersion is anomalous. $(iii)$ For $\ell = h$ or $(\ell,h)\rightarrow 0$ (i.e., no material microstructure), the wave velocities degenerate into the non-dispersive velocities of classical elastodynamics \cite{gourgiotis2013reflection, gourgiotis2015torsional, zisis2023wave}.

We finally mention that estimates for the two characteristic microstructural lengths $\ell$ and $h$ were given by Georgiadis \textit{et al.} \cite{georgiadis2004dispersive}, through a comparison between the Rayleigh-wave dispersion curves predicted by strain gradient theory and those obtained from a discrete particle theory based on an atomic-lattice approach. A more recent discussion of the values of these characteristic lengths was provided by Vavva \textit{et al.} \cite{vavva2009velocity}, in the context of cortical bone, where the length scales were related to the physical size of osteons. From an atomistic perspective, Maranganti \& Sharma \cite{maranganti2007novel} developed a molecular dynamics framework leveraging statistical mechanics to explicitly calculate and tabulate these internal strain gradient length parameters across diverse material systems, including metals, polymers, and semiconductors. Additionally, the role of these characteristic microstructural scales in capturing localized size effects and gradient-dominated couplings has been fundamentally explored by Sharma \textit{et al.} \cite{sharma2007possibility} within the mechanics of heterogeneous nanocomposites.

\section{Time-harmonic Green's function in plane-strain gradient elasticity}
\label{sec3}		
\noindent
In this section, we derive the Green's function for an infinite medium governed by strain gradient elasticity, subjected to a concentrated body force with time-harmonic variation under plane-strain conditions. The full analytical derivation is presented in \hyperref[A]{Appendix~A}, where the displacement equations of motion are solved using the double Fourier transform.

For an infinite body occupying the $(x,y)$-plane, with the $z$-axis normal to this plane, and under plane-strain time-harmonic conditions, the displacement and body force fields take the following form:
\begin{equation}\label{eq:3.5}
u_x(x,y,t) = u_x(x,y) e^{-i \omega t}, \quad u_y(x,y,t) = u_y(x,y) e^{-i \omega t}, \quad u_z \equiv 0,
\end{equation}
\begin{equation}\label{eq:3.6}
F_x(x,y,t) = F_x(x,y) e^{-i \omega t}, \quad F_y(x,y,t) = F_y(x,y) e^{-i \omega t}, \quad F_z \equiv 0.
\end{equation}
In view of Eqs.~\eqref{eq:3.5} and \eqref{eq:3.6}, the displacement equations of motion under time-harmonic conditions assume the form:
\begin{equation}\label{eq:3.7}
\left( 1 - \ell^2 \nabla^2 \right) \left[ \left( \lambda + 2\mu \right) \partial^2_x u_x + ( \lambda + \mu ) \partial_x \partial_y u_y + \mu \partial^2_y u_x \right] 
+ \rho \omega^2 \left( 1 - h^2 \nabla^2 \right) u_x + F_x = 0,
\end{equation}
\begin{equation}\label{eq:3.8}
\left( 1 - \ell^2 \nabla^2 \right) \left[ \left( \lambda + 2\mu \right) \partial^2_y u_y + ( \lambda + \mu ) \partial_x \partial_y u_x + \mu \partial^2_x u_y \right] 
+ \rho \omega^2 \left( 1 - h^2 \nabla^2 \right) u_y + F_y = 0,
\end{equation}
where $\nabla^2(\cdot) = \partial_x^2(\cdot) + \partial_y^2(\cdot)$.

To construct the Green's tensor, we consider a unit concentrated force acting successively along each coordinate direction. For a unit force applied in the $j$-direction, the $i^{\text{th}}$ body-force vector is written as $F_i^{(j)}=\delta_{ij}\delta(\mathbf{r}-\mathbf{r'})$ with $(i,j)=(x,y)$ and $\delta(\mathbf{r}-\mathbf{r'})$ being the two-dimensional Dirac delta distribution. The resulting displacement in the $i$-direction defines the Green's function component $u_i^{(j)} \equiv g_{ij}(\mathbf{r};\mathbf{r'})$. Thus, a unit force in the $x$-direction gives the first column of the Green's tensor, while a unit force in the $y$-direction gives the second column. The detailed derivation is provided in \hyperref[A]{Appendix~A}.

The displacement components of the Green's tensor are given by
\begin{equation}\label{eq:3.9}
g_{xx}
=
\frac{1}{4\pi\mu\beta^2}
\left[
\mathcal{R}_0(s)
-
\frac{(x-x')^2-(y-y')^2}{s^2}\mathcal{R}_2(s)
\right],
\end{equation}
\begin{equation}\label{eq:3.10}
g_{xy} =g_{yx}= - \frac{1}{2 \pi \mu \beta^2} \frac{\left( x - x' \right) \left( y - y' \right)}{s^2} \mathcal{R}_2(s),
\end{equation}
\begin{equation}\label{eq:3.11}
g_{yy} = \frac{1}{4 \pi \mu \beta^2} \left[ \mathcal{R}_0(s) + \frac{\left( x - x' \right)^2 - \left( y - y' \right)^2}{s^2} \mathcal{R}_2(s) \right],
\end{equation}
where $g_{ij}$ denotes the displacement in the $i$-direction due to a unit force in the $j$-direction, 
$\mathbf{s}=\mathbf{r}-\mathbf{r}'$, $s=|\mathbf{s}|$, and
\begin{equation}\label{eq:PhiPsi}
\mathcal{R}_0(s)=\Psi_P(s)+\beta^2\Psi_S(s),
\qquad
\mathcal{R}_2(s)=\Phi_P(s)-\beta^2\Phi_S(s).
\end{equation}
The auxiliary functions are defined as
\begin{equation}\label{eq:3.12}
\Psi_j(s) = \frac{1}{\Delta_j} \left[ \frac{\pi i}{2} H_0^{(1)}(q_{1j} s) - K_0(q_{2j} s) \right], \quad j = P, S,
\end{equation}
\begin{equation}\label{eq:3.13}
\Phi_j(s) = \frac{1}{\Delta_j} \left[ \frac{\pi i}{2} H_2^{(1)}(q_{1j} s) + K_2(q_{2j} s) \right], \quad j = P, S,
\end{equation}
\begin{equation}\label{eq:3.14}
\beta = \frac{c_P}{c_S} = \frac{k_S}{k_P} = \sqrt{\frac{2 \left( 1 - \nu \right)}{1 - 2 \nu}}, \quad k_j = \frac{\omega}{c_j}, \quad j = P, S,
\end{equation}
where $H_n^{(1)}$ and $K_n$ denote the Hankel function of the first kind and the modified Bessel function of the second kind of order $n$, respectively.

The quantities $q_{1j}$ and $q_{2j}$ are real wavenumbers and are associated with the roots of the dispersion relations of the pressure and shear waves given in Eqs.~\eqref{eq:2.14}, \eqref{eq:2.15}. Specifically, $\pm q_{1j}$ correspond to the real roots of the dispersion relations and are associated with propagating waves, whereas $ \pm i q_{2j}$ correspond to purely imaginary roots and are associated with evanescent modes
\begin{equation}\label{eq:3.15}
q_{1j} \equiv q_{1j}(\omega) = \left[ \frac{\Delta_j - \left( 1 - h^2 k_j^2 \right)}{2 \ell^2} \right]^{1/2} > 0, \quad j = P, S,
\end{equation}
\begin{equation}\label{eq:3.16}
q_{2j} \equiv q_{2j}(\omega) = \left[ \frac{\Delta_j + \left( 1 - h^2 k_j^2 \right)}{2 \ell^2} \right]^{1/2} > 0, \quad j = P, S,
\end{equation}
\begin{equation}\label{eq:3.17}
\Delta_j \equiv \Delta_j(\omega) = \sqrt{\left( 1 - h^2 k_j^2 \right)^2 + 4 \ell^2 k_j^2} > 0, \quad j = P, S.
\end{equation}
The positive values of $q_{1j}$ and $q_{2j}$ are chosen so that the Green's function satisfies the appropriate radiation and decay conditions: with the time-harmonic convention $e^{-i\omega t}$, the terms involving $H_n^{(1)}(q_{1j}s)$ represent outgoing cylindrical waves, whereas the terms involving $K_n(q_{2j}s)$ describe exponentially decaying evanescent modes as $s$ increases.

The Green's tensor can be compactly expressed in matrix form as
\begin{equation}\label{eq:3.18}
\mathbf{g}(\mathbf{r};\mathbf{r}')
=
\frac{1}{4\pi\mu\beta^2}
\Bigl\{
\left[\mathcal{R}_0(s)+\mathcal{R}_2(s)\right]\mathbf{I}
-
2\mathcal{R}_2(s)\,\hat{\mathbf{s}}\otimes\hat{\mathbf{s}}
\Bigr\}.
\end{equation}
where $s=|\mathbf{s}|$ denotes the magnitude of $\mathbf{s}$, and $\mathbf{\hat{s}}=\mathbf{s}/s$ is the unit vector in the direction of $\mathbf{s}$. Visualizations for the Green's function are provided in \hyperref[A]{Appendix~A}. 

Unlike the corresponding solution of classical elasticity given in \cite{eason1956generation}, Eq.~\eqref{eq:3.18} is bounded in the limit where $ \mathbf{r} \rightarrow \mathbf{r}' $, (i.e., as $s \rightarrow 0$), indicating that the logarithmic singularity of classical elasticity is now eliminated. In fact, as $s \rightarrow 0$, we obtain
\begin{equation}\label{eq:3.19}
\mathbf{g}(\mathbf{r'}; \mathbf{r'}) = \frac{1}{4 \mu} \left\lbrace \frac{1}{\pi} \left[ \frac{1}{\Delta_S} \ln \left( \frac{q_{2S}}{q_{1S}} \right) + \frac{1}{\beta^2 \Delta_P} \ln \left( \frac{q_{2P}}{q_{1P}} \right) \right] + \frac{i}{2} \left[ \frac{1}{\Delta_S} + \frac{1}{\beta^2 \Delta_P} \right] \right\rbrace \mathbf{I},
\end{equation}
and
\begin{equation}\label{Green derivatives}
\frac{d \mathbf{g}}{d s} = \mathcal{O}(s \ln s) \mathbf{I}, \qquad \frac{d^2 \mathbf{g}}{d s^2} = \mathcal{O}(\ln s) \mathbf{I}, \qquad \text{as } s \to 0.
\end{equation}
It follows that the displacement Green's tensor is $C^1$-continuous at the source. Consequently, a time-harmonic concentrated in-plane force induces bounded strains $\varepsilon_{ij}$, whereas the strain gradients $\kappa_{ijk}$ exhibit a logarithmic singularity at $s=0$. Nevertheless, the total strain energy remains finite in any bounded region containing the source, illustrating the regularizing effect of strain gradient elasticity. This is in marked contrast with classical elasticity, where the strain energy becomes singular and unbounded in the 2D time-harmonic plane-strain Kelvin problem.

Finally, we note that the corresponding Green's tensor of classical elastodynamics is recovered in the limit $(\ell,h)\to(0,0)$. Similarly, the static strain gradient Green's tensor given in \cite{polyzos2003boundary, gourgiotis2018concentrated} is obtained by taking the limit $\omega\to0$.

\section{Scattering of P and SV waves by point scatterers}
\label{sec4}
\begin{figure}[!htb]
\centering
\includegraphics[scale=0.70]{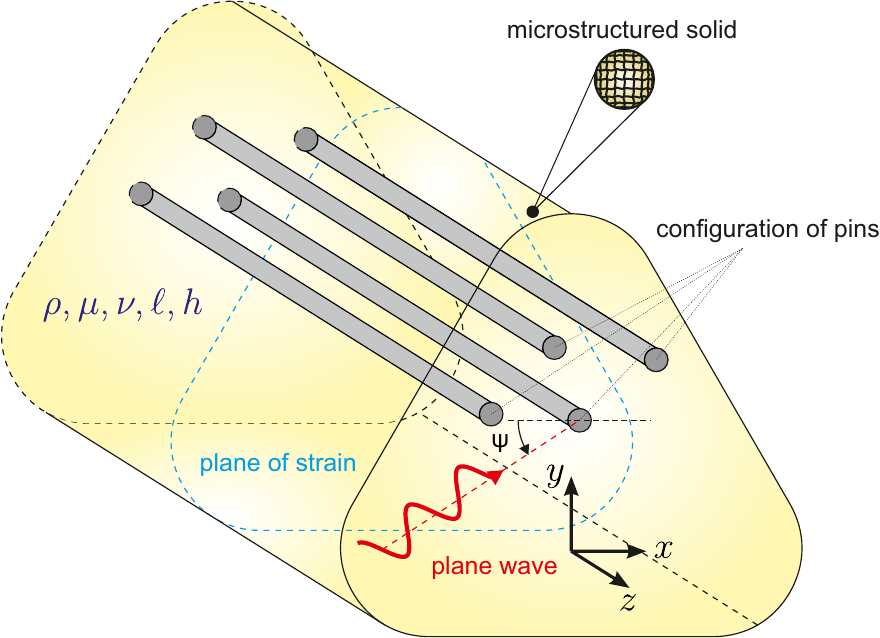}
\caption{Configuration of pins embedded in an infinite elastic microstructured medium subjected to an incident harmonic plane-wave, under plane-strain conditions.}
\label{fig01}
\end{figure}

\noindent
The problem under investigation is illustrated in Fig.~\ref{fig01}, which shows plane waves propagating through an infinite medium containing a cluster of point scatterers, i.e., pins. In the plane-strain setting, the pins are effectively rigid, fibre-like inclusions with zero cross-sectional area along the $z$-axis. An incident time-harmonic wave of unit amplitude propagates in the $(x,y)$-plane, generating an in-plane displacement of the form
\begin{equation}\label{eq:4.1}
\mathbf{u}^{in}(x,y,t) = \mathbf{U}^{in}(\mathbf{r}) e^{-i \omega t},
\end{equation}
where $\omega$ is the frequency of the incident excitation and $\mathbf{r} = (x,y) = (r \cos(\theta), r \sin(\theta))$, with $\theta$ denoting the polar angle.

It is important to note that the incident plane wave may comprise a combination of both pressure (P) and vertically polarized shear (SV) wave components. In such cases, the incident wave field can be represented as the superposition of two unit amplitude incident wave fields: one corresponding to a P-wave and the other to an SV-wave.

A prescribed incident plane wave may be taken as a P-wave, an SV-wave, or a linear combination of the two. In the latter case, the incident displacement field can be written as
\begin{equation}
\label{eq:4.2}
\mathbf{U}^{in}(\mathbf{r}) =
B_P \, \mathbf{d}_P e^{i q_{_{1P}}\, r\cos(\theta-\psi)}
+ B_S \, \mathbf{d}_S e^{i q_{_{1S}}\, r\cos(\theta-\psi)},
\end{equation}
where
\begin{equation}
\mathbf{d}_P=\cos\psi\,\hat{\mathbf{x}}+\sin\psi\,\hat{\mathbf{y}},
\qquad
\mathbf{d}_S=-\sin\psi\,\hat{\mathbf{x}}+\cos\psi\,\hat{\mathbf{y}}.
\end{equation}
Here, $q_{1P}$ and $q_{1S}$ are the positive real roots of the P and SV-wave dispersion relations, respectively, as defined by Eq.~\eqref{eq:3.15} for $j=P,S$. The coefficients $B_P$ and $B_S$ prescribe the amplitudes of the incident P and SV components. Moreover, $\psi$ is the incidence angle measured from the positive $x$-axis to the direction of wave propagation, while $\hat{\mathbf{x}}$ and $\hat{\mathbf{y}}$ are unit vectors along the positive $x$- and $y$-axes, respectively. For a purely incident P-wave, $B_S=0$, and the particle motion is parallel to the direction of propagation. For a purely incident SV-wave, $B_P=0$, and the particle motion is perpendicular to the direction of propagation while remaining within the plane.

To facilitate the numerical computations, the Green's function \eqref{eq:3.18} is expressed using the following dimensionless parameters:
\begin{equation}\label{eq:4.5}
X = \frac{x}{\ell}, \quad Y = \frac{y}{\ell}, \quad H = \frac{h}{\ell}, \quad  \Omega = \frac{\ell \omega}{c_P} .
\end{equation}
This normalisation facilitates the classification of the dispersion behavior using solely the dimensionless parameter $H$ as explained in \hyperref[sec2]{Section~2}.

We consider now a cluster of $N_p$ pins at points with position vectors $\mathbf{R}_n$ = $(X_n,Y_n)$. The total non-dimensional displacement $\mathbf{U}(\mathbf{R})$ (i.e., normalized with the amplitude of the incident plane wave) at any point of the infinite domain with position vector $\mathbf{R} = (X,Y)$, can be expressed as the superposition of the displacement generated by the incident wave, given by Eq.~\eqref{eq:4.2}, and the displacement induced by the scattering of the plane waves due to the presence of all the pins. We enforce the null-displacement constraint at the pins by introducing fictitious concentrated body forces at the pin locations. These forces are chosen so that the displacements they generate exactly cancel the displacements produced by the incident plane waves at the pin locations. By the principle of superposition, the total displacement at each pin is therefore zero. The displacements generated by the concentrated body forces are readily determined using the Green's function \eqref{eq:3.18}. Thus, the total displacement is given by superposition as
\begin{equation}\label{eq:4.6}
\mathbf{U}(\mathbf{R}) = \mathbf{U}^{in}(\mathbf{R}) + \sum_{n=1}^{N_p} \left[ \mathbf{g}(\mathbf{R}; \mathbf{R}_n) \cdot \mathbf{A}_n \right],
\end{equation}
where $\mathbf{A}_n = (A_n^x, A_n^y)$ ($n = 1,2,...,N_p$) are constant vectors which are obtained by solving the following non-homogeneous system of $N_p$ vector equations at the pin locations $\mathbf{R} = \mathbf{R}_m$ ($m = 1,2,...,N_p$) where the displacement vector becomes zero:
\begin{equation}\label{eq:4.7}
\mathbf{0} = \mathbf{U}^{in}(\mathbf{R}_m) + \sum_{n=1}^{N_p} \left[ \mathbf{g}(\mathbf{R}_m; \mathbf{R}_n) \cdot \mathbf{A}_n \right], \quad m = 1,2,...,N_p.
\end{equation}
The solution of the system of equations \eqref{eq:4.7} can be expressed as the following matrix equation
\begin{equation}\label{eq:4.8}
\tilde{\mathbf{A}} = - \mathbf{G}^{-1} \cdot \tilde{\mathbf{U}}^{in},
\end{equation}
where $\tilde{\mathbf{A}}$ is a $2N_p \times 1$ column vector, $\mathbf{G}$ is the $2N_p \times 2N_p$ non-dimensional Green's matrix with block components $\mathbf{G}_{mn}(\mathbf{R}_m;\mathbf{R}_n)$, and $\tilde{\mathbf{U}}^{in}$ is a $2N_p \times 1$ column vector containing the non-dimensional displacement components of the prescribed incident field evaluated at the pin locations in the absence of pins. Since the time-harmonic Green's tensor is generally complex-valued, $\mathbf{G}$ is a complex symmetric matrix; reciprocity implies the block symmetry $\mathbf{G}_{mn}=\mathbf{G}_{nm}^{T}$. Extensions of the displacement reciprocity theorem to strain gradient elasticity are provided in \cite{giannakopoulos2006reciprocity, georgiadis2006energy}
\begin{equation}\label{eq:blocks}
\tilde{\mathbf{A}} =
\left[
\begin{array}{c}
A_1^x \\[6pt]
A_1^y \\[6pt]
A_2^x \\[6pt]
A_2^y \\[6pt]
\vdots \\[6pt]
A_{N_p}^x \\[6pt]
A_{N_p}^y
\end{array}
\right],
\quad
\tilde{\mathbf{U}}^{in} =
\left[
\begin{array}{c}
U_x^{in}(\mathbf{R}_1) \\[6pt]
U_y^{in}(\mathbf{R}_1) \\[6pt]
U_x^{in}(\mathbf{R}_2) \\[6pt]
U_y^{in}(\mathbf{R}_2) \\[6pt]
\vdots \\[6pt]
U_x^{in}(\mathbf{R}_{N_p}) \\[6pt]
U_y^{in}(\mathbf{R}_{N_p})
\end{array}
\right],
\quad
\mathbf{G}_{mn} =
\begin{bmatrix}
g_{xx}(\mathbf{R}_m; \mathbf{R}_n) & g_{xy}(\mathbf{R}_m; \mathbf{R}_n) \\
g_{yx}(\mathbf{R}_m; \mathbf{R}_n) & g_{yy}(\mathbf{R}_m; \mathbf{R}_n)
\end{bmatrix}.
\end{equation}
The diagonal blocks $\mathbf{G}_{nn}$ are well defined because the strain gradient Green's tensor remains bounded at the source; they are evaluated using the limiting expression given in Eq.~\eqref{eq:3.19}.

The constant vectors $\mathbf{A}_n$ contained in $\tilde{\mathbf{A}}$ represent the normalized reaction forces exerted by the pins and determine the amplitude of the scattered field generated by each point constraint. Through the superposition formula in Eq.~\eqref{eq:4.6}, these constants influence the displacement field throughout the infinite domain. Consequently, large values of $\mathbf{A}_n$ are associated with pronounced scattered-field amplitudes and may lead to strong displacement responses in the medium. According to Eq.~\eqref{eq:4.8}, $\tilde{\mathbf{A}}$ depends on the Green's matrix $\mathbf{G}$ and on the vector $\tilde{\mathbf{U}}^{in}$, which contains the values of the prescribed incident displacement field evaluated at the pin locations. Thus, for a fixed pin configuration and a prescribed incident wave, the scattered-field amplitudes are determined by the combined effect of the Green's matrix and the sampled incident field.

For a fixed pin configuration and given mechanical parameters $(\nu,\ell,h)$, the Green's matrix $\mathbf{G}$ depends on the frequency $\omega$ through the dimensionless parameter $\Omega$. The determinant $\left\lvert \det(\mathbf{G}) \right\rvert$ is therefore used as a scalar diagnostic of the frequency-dependent behavior of the algebraic system. Since $\log\left(\left\lvert \det(\mathbf{G}) \right\rvert\right)$ may exhibit an overall monotonic trend with increasing $\Omega$, candidate resonant frequencies are not identified from its absolute value, but from pronounced local minima relative to the surrounding frequency trend.

Using the logarithmic measure
\begin{equation}\label{eq.Gamma}
\Gamma = \log\left(\left\lvert \det(\mathbf{G}) \right\rvert\right),
\end{equation}
which enhances the contrast between sharp and shallow minima, candidate resonant frequencies are first identified from the local-minimum conditions
\begin{equation}\label{eq:4.9}
\frac{d \Gamma}{d \Omega}=0,
\qquad
\frac{d^{2}\Gamma}{d\Omega^{2}}>0 .
\end{equation}
However, these conditions alone are not sufficient to distinguish trapping modes from ordinary non-localized scattering responses. In the computations reported below, trapping modes are associated with sharp, well-developed minima of $\Gamma$. Such minima are characterized by a pronounced local maximum of the curvature $d^{2}\Gamma/d\Omega^{2}$. Accordingly, among the local minima satisfying Eq.~\eqref{eq:4.9}, those retained as trapping resonances are selected by requiring that $d^{3}\Gamma/d\Omega^{3}$ be sufficiently small, within the adopted numerical tolerance, and that $d^{4}\Gamma/d\Omega^{4}<0$. These higher-order derivative conditions are therefore not introduced as exact analytical resonance conditions, but as numerical diagnostic criteria for selecting the sharp minima associated with localized trapping modes. To compute these quantities accurately, the Jacobi formula is employed for the derivatives of $\det(\mathbf{G})$.

When the additional condition on $d^{3}\Gamma/d\Omega^{3}$ is satisfied to numerical precision, the corresponding trapping mode approaches a nearly perfect localization regime, in which the motion is confined almost entirely within the pin configuration. This behavior is observed in Figs.~\ref{fig06}e, \ref{fig12}, and \ref{fig14}c. Such strong localization is rare for circular arrangements consisting of a single ring of pins, but may arise in configurations with multiple corner-like features, as shown in Fig.~\ref{fig12}; see also \cite{alevras2026scattering}.

It is important to emphasize that the term ``resonance'' is used here in a generalized sense. Unlike classical eigenvalue problems, where resonance is associated with non-trivial solutions of a homogeneous system, the present formulation leads to a non-homogeneous algebraic system. Here, resonance refers to frequencies at which the prescribed incident field, through its interaction with the pin configuration, produces a strongly amplified scattered response. Trapping resonances form a more restrictive class of such responses, characterized by localization of the displacement field within the pinned region.

\begin{figure}[!htb]
\centering
\includegraphics[scale=0.45]{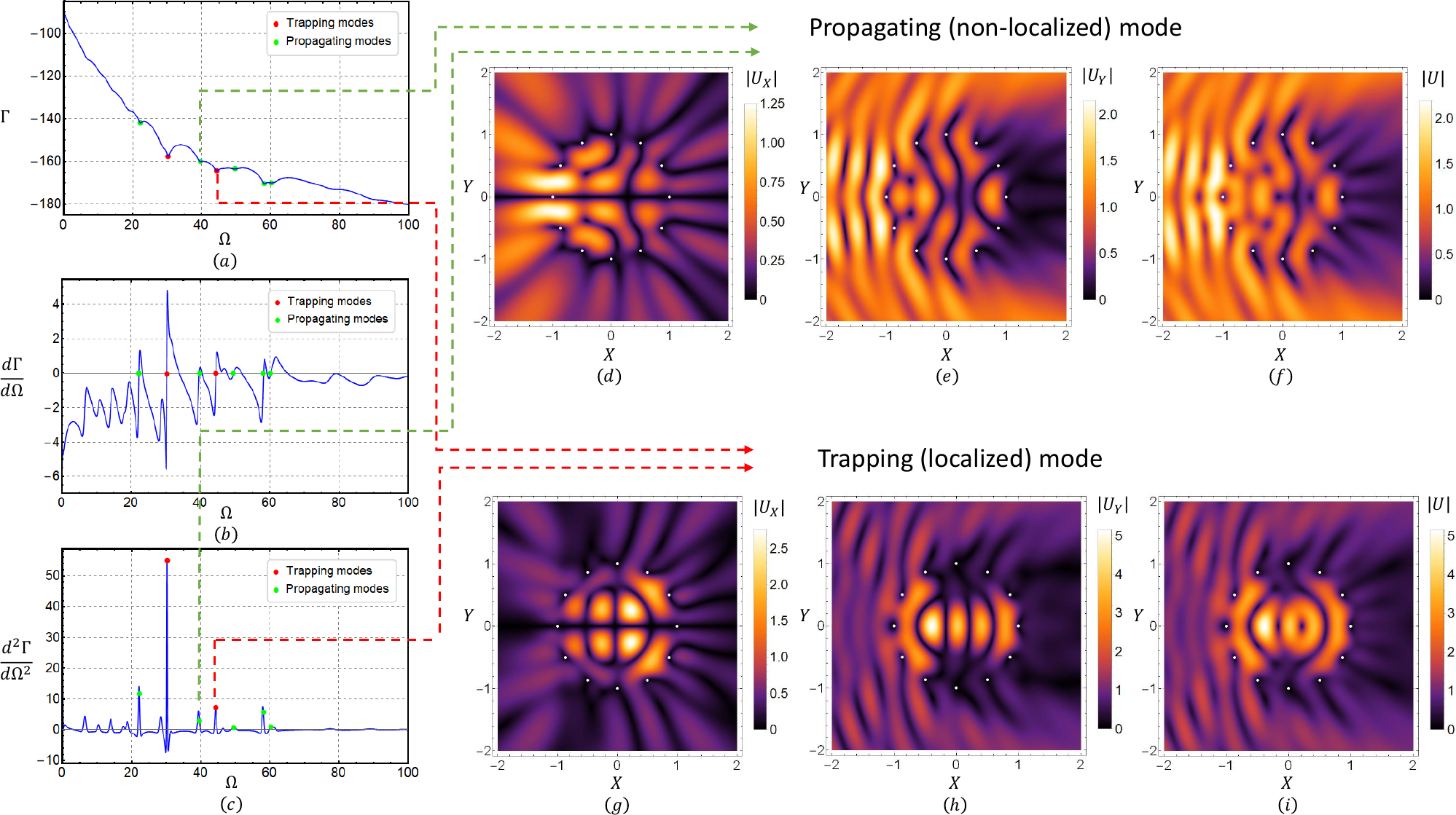}
\caption{$(a)-(c)$ detected resonant frequencies, $(d)-(f)$ propagating (non-localized) mode $\Omega = 39.63393$, and $(g)-(i)$ trapping (localized) mode $\Omega = 44.41294$, for a circular configuration, subjected to an incident SV-wave of unit amplitude. In all cases, $H=0$, $\nu=0.25$, $\psi=0^{\circ}$, $N_p = 12$.}
\label{fig02}
\end{figure}

Figs.~\ref{fig02}a-c illustrate the detected local minima for a circular arrangement of $12$ pins, subjected to unit amplitude SV waves, for values of $\Omega < 100$. In general, sharper local minima of $\lvert \det(\mathbf{G}) \rvert$ (or equivalently of $\Gamma$) correspond to resonant modes that produce greater displacements in the body. These sharp minima are characterized by a peak in the second derivative of $\Gamma$ with respect to $\Omega$ and satisfy the additional higher-order diagnostic criteria described above. In such cases, the wave disturbance remains predominantly confined within the pin cluster. This behavior defines the trapping (localized) modes. In contrast, for other local minima of $\lvert \det(\mathbf{G}) \rvert$ (marked in green), the disturbance does not localize; instead, plane waves propagate through the pin array, producing larger displacements outside the cluster. A similar non-localized response is also observed at frequencies that do not satisfy the trapping criteria. This correlation between the sharpness of local minima and the nature of resonant modes holds consistently across all pin configurations considered in this study.

Figures \ref{fig02}d-i present the distribution of the absolute value of the non-dimensional displacement components, $\lvert U_X \rvert$ and $\lvert U_Y \rvert$, as well as the magnitude of the non-dimensional displacement vector, $\lvert U \rvert$, for a circular array of unit radius consisting of $12$ pins. The configuration is subjected to an incident SV-wave of unit amplitude propagating in the positive $X-$direction $(\psi=0)$, under different types of resonances. Figs.~\ref{fig02}d-f display the formation of distinct propagation paths extending beyond the pin array. This corresponds to the $3^{rd}$ local minimum point in Fig.~\ref{fig02}a, which does not correspond to a local maximum in $d^2 \Gamma / d\Omega^2$, suggesting a non-localized propagating response rather than a trapping-dominated resonant response. In contrast, in Figs.~\ref{fig02}g-i the motion is mainly confined within the configuration of pins. This corresponds to the $4^{th}$ local minimum point in Fig.~\ref{fig02}a, which aligns with a local maximum point of $d^2 \Gamma /d\Omega^2$ as shown in Fig.~\ref{fig02}c, indicating a trapping mode. The analysis further indicates that maximum absolute displacements are consistently higher for trapping modes compared to these propagating responses, underscoring the greater ability of the former to localize energy within the pin configuration.

It is important to note that Figures \ref{fig02}a-c are unaffected by the type of incident plane wave. This is because the determinant of the Green's matrix does not depend on whether the incoming wave is of type P or SV.

\begin{figure}[!htb]
\centering
\includegraphics[scale=0.45]{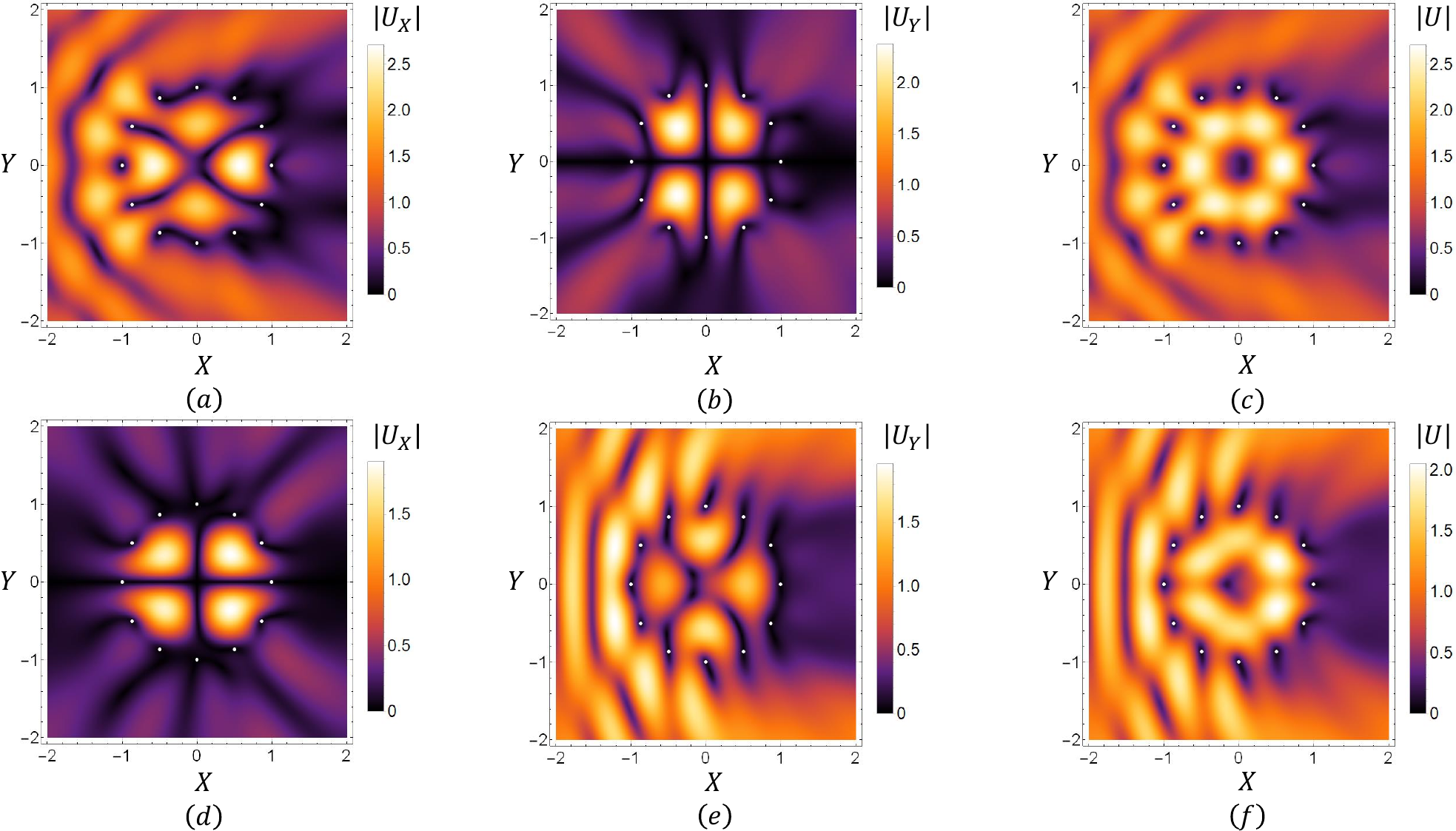}
\caption{Distribution of the non-dimensional displacement components and the magnitude of the non-dimensional displacement vector, for a circular configuration, subjected to an incident $(a)-(c)$ P and $(d)-(f)$ SV-wave of unit amplitude. In all cases, $H=0$, $\nu=0.25$, $\psi=0^{\circ}$, $\Omega=22.22847$, $N_p = 12$.}
\label{fig03}
\end{figure}

The results also show that, in certain cases, one displacement component may exhibit partial localization within the pin configuration, even when the full trapping criteria are not satisfied. In all such cases, the localized displacement component is oriented perpendicular to the polarization direction, regardless of the wave type (P or SV). This behavior is illustrated in Fig.~\ref{fig03}, which show the distribution of the non-dimensional displacements at the first resonant mode.

In Figs.~\ref{fig03}a–c, a P-wave propagates along the positive $X$-axis with particle motion polarized in the $X$-direction; here, localization occurs in the $Y$-direction. In contrast, in Figs.~\ref{fig03}d–f, an SV-wave propagates along the positive $X$-axis with polarization in the $Y$-direction, and localization arises in the $X$-direction.

\section{Circular configurations of pins}
\label{sec5}
\noindent
We first focus on a circular point-scatterer configuration in which the pins are distributed along the perimeter of a circle with unit non-dimensional radius, $a/\ell=1$. The system response is examined as a function of the parameters $(H,\nu,\psi,\Omega,N_p)$. Extensions involving multiple circular arrangements, including displaced and offset configurations, are considered in \hyperref[sec5.5]{Section~5.5}.

\subsection{The effect of the incident plane wave type}
\noindent
With respect to the type of incident plane wave, namely P, SV, or a linear combination of the two, the response of circular pin clusters cannot be classified solely on the basis of the incident-wave polarization. Scattering at the pins induces mode conversion, so that both P and SV components are generally present in the total field. Consequently, the wave pattern and the degree of localization are governed by the combined effect of the incident polarization, the angle of incidence, the material parameters, and the pin geometry.

This point is illustrated in Fig.~\ref{fig03}, where the first resonant mode of the circular configuration is shown for incident P and SV waves. Although the determinant-based resonance frequency is the same for both incident-wave types, the displacement components excited by the incident field differ markedly. For P-wave incidence along the positive $X$-axis, the imposed particle motion is primarily in the $X$-direction, yet the localized response appears mainly in the transverse $Y$-component. Conversely, for SV-wave incidence, where the imposed particle motion is in the $Y$-direction, the localized response appears mainly in the $X$-component. Thus, Fig.~\ref{fig03} shows that partial localization may occur in the displacement component perpendicular to the incident polarization, even when the full displacement magnitude does not exhibit a trapping-dominated response. This behavior reflects the vectorial nature of the plane-strain scattering problem and the mode conversion induced by the pin constraints.

\subsection{The effect of dispersion type}
\noindent
The analysis shows that the type of dispersion, characterized by the non-dimensional microstructural ratio $H=h/\ell$, has a decisive influence on the behavior of $\Gamma=\log\left(\left\lvert\det(\mathbf{G})\right\rvert\right)$ and, consequently, on the system response. For the configurations examined, sharp local minima associated with trapping modes are observed primarily in the anomalous dispersion regime, $H<1$. This indicates that localization is strongly promoted when the energetic length scale $\ell$ dominates over the microinertial length scale $h$.

\begin{figure}[htbp]
    \centering
    \includegraphics[width=\textwidth]{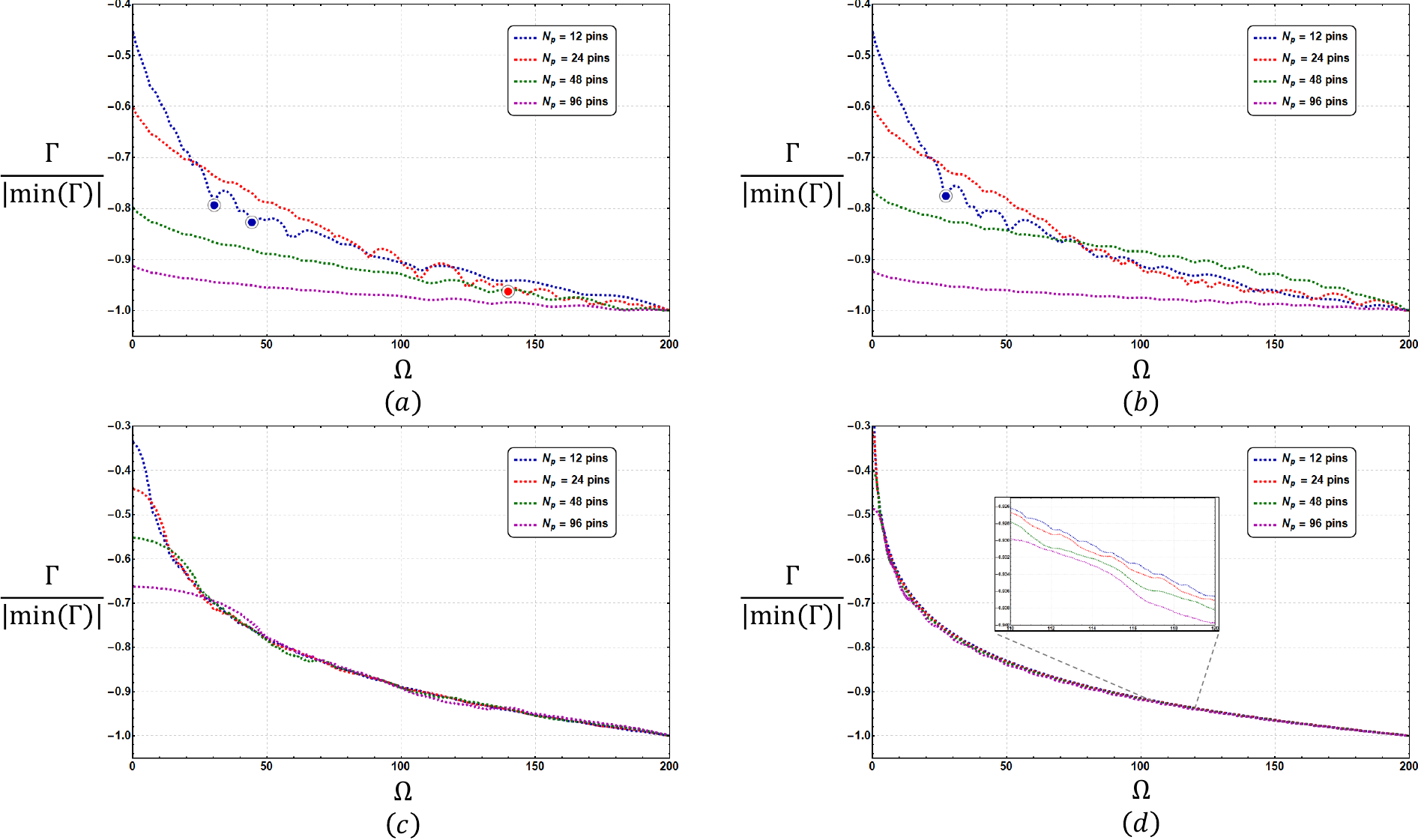}
    \caption{Variation of the normalized logarithmic determinant $\Gamma/ |\min(\Gamma)|$ and detected trapping (localized) modes for circular pinned configurations in the low-frequency range $0<\Omega<200$: (a) $H=0$, (b) $H=0.1$, (c) $H=1$, and (d) $H=10$. In all cases, $\nu=0.25$.}
    \label{fig04}
\end{figure}

\begin{figure}[htbp]
    \centering
    \includegraphics[width=\textwidth]{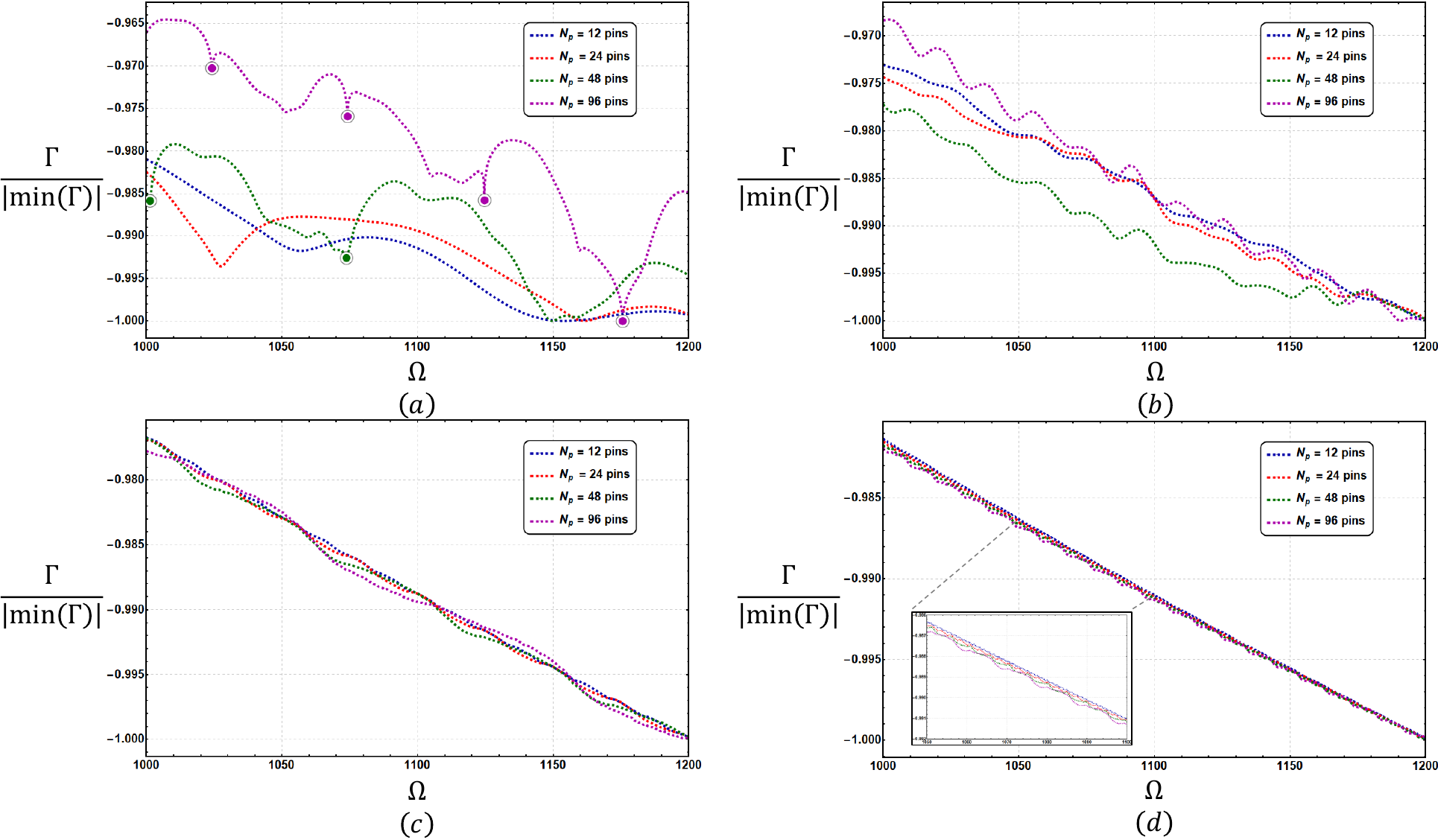}
    \caption{Variation of the normalized logarithmic determinant $\Gamma/|\min(\Gamma)|$ and detected trapping (localized) modes for circular pinned configurations in the high-frequency range $1000<\Omega<1200$: (a) $H=0$, (b) $H=0.1$, (c) $H=1$, and (d) $H=10$. In all cases, $\nu=0.25$.}
    \label{fig05}
\end{figure}

As shown in Figs.~\ref{fig04} and \ref{fig05}, for circular configurations of pins, distinct sharp local minima become especially pronounced for $H<1$, and in particular for $H=0$, corresponding to zero microinertia. These minima satisfy the local-minimum condition in Eq.~\eqref{eq:4.9}, together with the higher-order (curvature-based) selection criteria described in \hyperref[sec4]{Section~4}. As $H$ increases, the number of local minima in $\Gamma$ may increase; however, these minima become less sharp and no longer necessarily correspond to strong resonant localization. Beyond a certain range, the response transitions towards a delocalized regime. Closer inspection of Figs.~\ref{fig04}d and \ref{fig05}d shows that, in the case of normal dispersion, $H>1$, the oscillatory behavior of $\Gamma$ is characterized by closely spaced local minima with attenuated curvature peaks. These features indicate a weakening of trapping resonances and a shift towards smoother, less localized dynamics. As shown below, in this regime a cloaking-type response becomes dominant, with the pins only weakly perturbing the incident field over the frequency range considered. Figs.~\ref{fig04} and \ref{fig05} are presented separately for visualization purposes, in order to enhance clarity and facilitate the identification of local minima and candidate resonant frequencies.


In the present model, normal dispersion occurs when the microinertial length dominates the energetic length scale, $h>\ell$, whereas anomalous dispersion occurs when $\ell>h$. The latter regime is commonly associated with media in which microstructural stiffness effects are sufficiently strong to produce enhanced dispersive behavior, as in certain architected materials and phononic systems \cite{srivastava2015elastic}. It is within this anomalous dispersion regime that the present computations reveal the most pronounced effects, including sharp resonant minima and strong wave localization.

As illustrated in Fig.~\ref{fig04}a, at low values of the non-dimensional frequency $\Omega$, circular configurations with fewer pins exhibit more resonances, whereas at higher values of $\Omega$, configurations with more pins resonate more frequently, as shown in Fig.~\ref{fig05}a. This trend reflects the role of the inter-pin spacing: at lower frequencies, the wavelength is comparable to the spacing in sparser configurations, while at higher frequencies it becomes comparable to the finer spacing of denser configurations.

\subsubsection{$H < 1$: Localized resonant regime}
\noindent
In the anomalous dispersion regime, $H<1$, the system exhibits sharp local minima of $\Gamma$ that are associated with localized trapping modes. These frequencies satisfy the local-minimum condition in Eq.~\eqref{eq:4.9}; the trapping modes are then identified by applying the higher-order curvature criteria described in \hyperref[sec4]{Section~4}. This regime is characterized by pronounced displacement amplification and strong confinement of the wave field within the pin configuration.

\begin{figure}[!htb]
\centering
\includegraphics[scale=0.45]{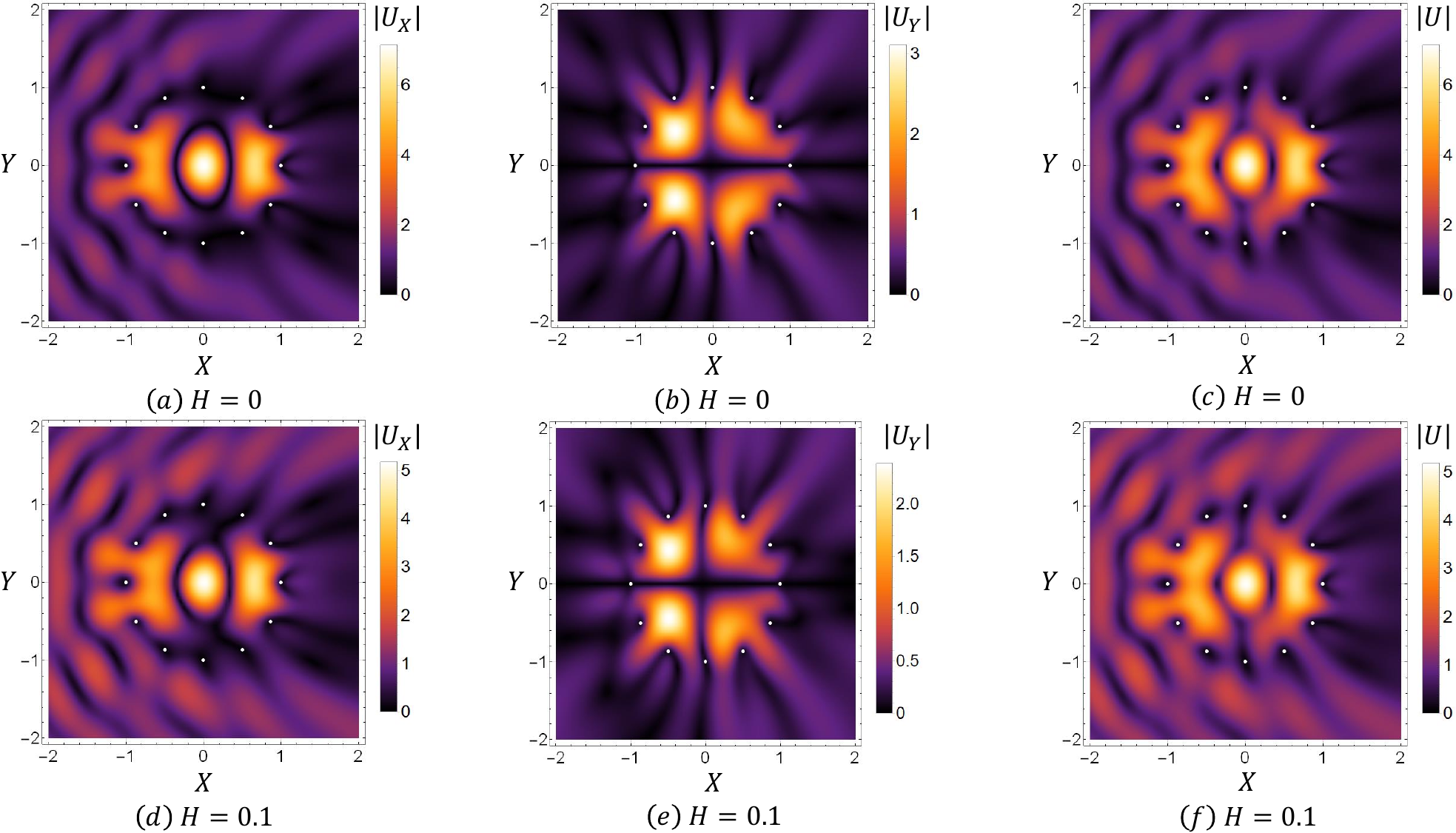}
\caption{Distribution of the non-dimensional displacement components and the magnitude of the non-dimensional displacement vector at two nearby trapping (localized) modes, for a circular configuration, subjected to an incident $P$-wave of unit amplitude. $(a)-(c)$ $\Omega=30.28867$ and $(d)-(f)$ $\Omega=26.86327$. In all cases, $\nu=0.25$, $\psi=0^{\circ}$, $N_p = 12$.}
\label{fig06}
\end{figure}

Figure~\ref{fig06} illustrates the response at two nearby trapping modes for different values of $H$. The comparison shows that decreasing $H$ strengthens the localization effect, with the case $H=0$ producing a more pronounced resonant response than the case $H=0.1$. This trend persists regardless of the number of pins $N_p$ or the Poisson's ratio $\nu$.

\begin{figure}[!htb]
\centering
\includegraphics[scale=0.45]{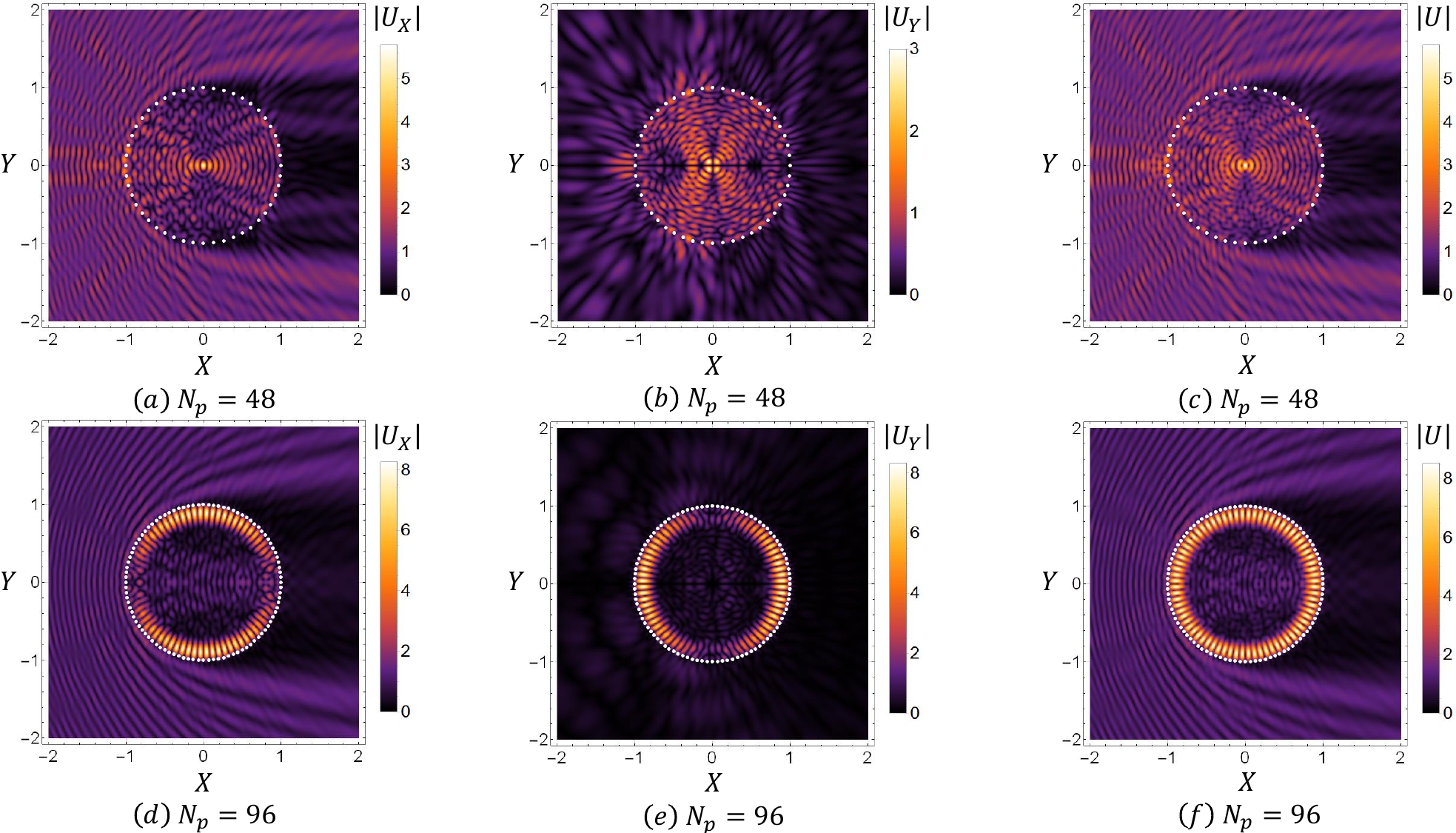}
\caption{Distribution of the non-dimensional displacement components and of the magnitude of the non-dimensional displacement vector at two nearby high-frequency trapping modes for circular configurations with different numbers of pins: (a)--(c) $N_p=48$, $\Omega=1073.76825695$; (d)--(f) $N_p=96$, $\Omega=1074.099918$. In all cases, $H=0$, $\nu=0.25$, $\psi=0^\circ$, and the incident wave is SV.}
\label{fig07}
\end{figure}

Moreover, within the frequency ranges examined, the strongest displacement localization is generally observed at larger values of the non-dimensional frequency $\Omega$ and for denser circular configurations, as shown in Fig.~\ref{fig07}. This behavior is consistent with the reduced spacing between neighbouring pins: at higher frequencies, the wavelength becomes comparable to the finer inter-pin spacing of denser configurations, promoting stronger multiple-scattering interactions and enhanced localization. In this case, the trapped field is concentrated mainly along the pinned perimeter rather than throughout the entire interior of the cluster. This indicates the formation of a perimeter-localized trapping mode. In this regime, the dense circular array behaves as an effective annular scattering barrier, supporting a response that is localized along the circular boundary and resembles a whispering-gallery-type mode.

\subsubsection{$H = 1$: Non-dispersive behavior}
\noindent
The case $H=1$, corresponding to $h=\ell$, represents the non-dispersive limit of the present strain-gradient model. Indeed, the dispersion relations reduce to $\omega=c_j q$, with $j=P,S$, so that the propagating P and SV waves travel with the classical phase velocities. At the same time, the medium does not reduce to classical elasticity, since the characteristic length $\ell$ remains finite and the Green's tensor retains the regularizing contribution associated with the evanescent roots. In particular, for $H=1$ one obtains $q_{1j}=k_j$ and $q_{2j}=1/\ell$.

\begin{figure}[!htb]
\centering
\includegraphics[scale=0.45]{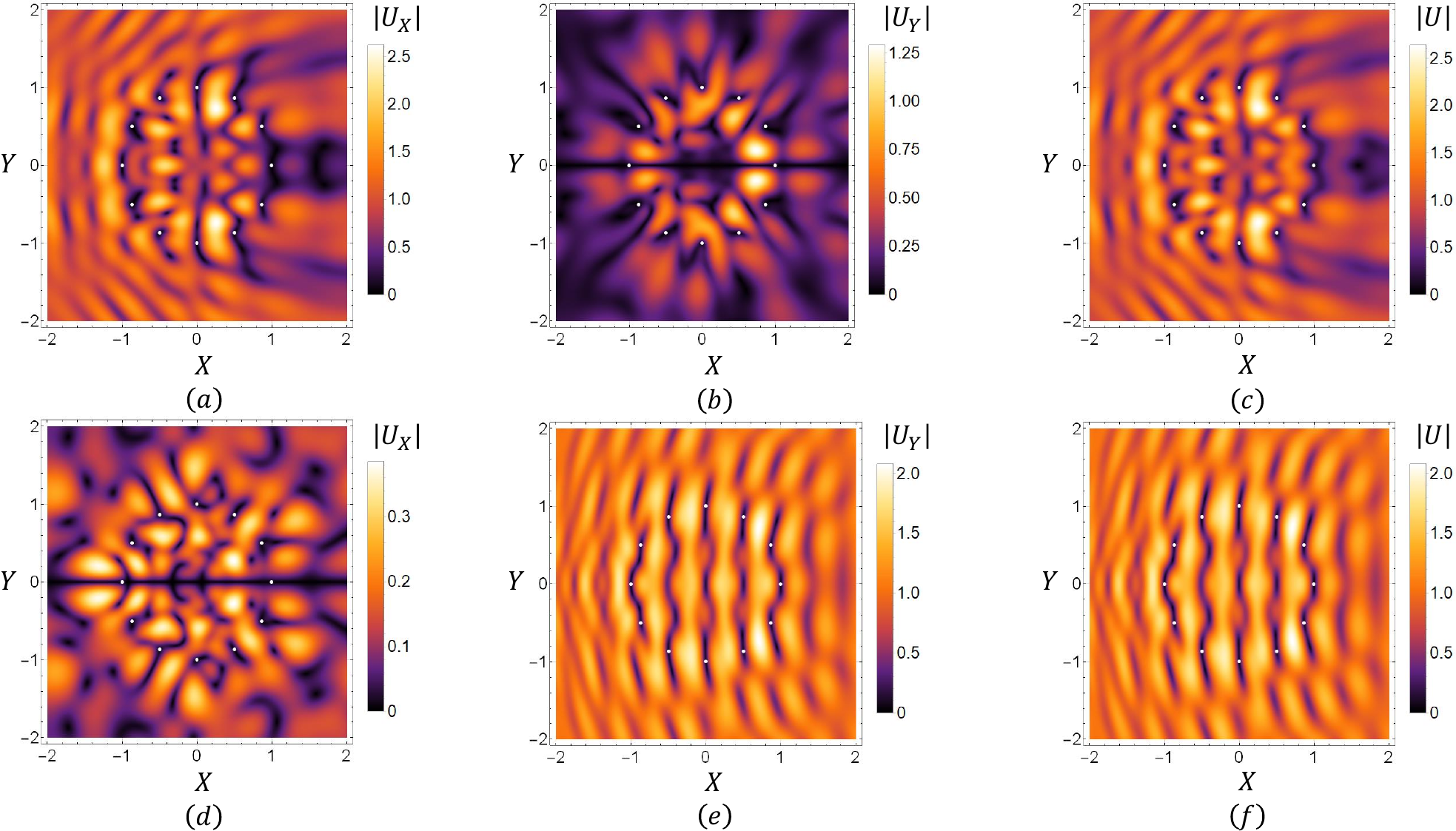}
\caption{Distribution of the non-dimensional displacement components and the magnitude of the non-dimensional displacement vector for a circular configuration, subjected to an incident $(a)-(c)$ P-wave and $(d)-(f)$ SV-wave of unit amplitude. In all cases, $H=1$, $\nu=0.25$, $\psi=0^{\circ}$, $\Omega=8.07281$, $N_p = 12$.}
\label{fig08}
\end{figure}

In this regime, the system response is more difficult to categorize. Since the propagating waves are non-dispersive, the sharp trapping resonances observed for $H<1$ are weakened. However, because the strain gradient regularization is still present, the response does not coincide with that of classical elastodynamics. As a result, the displacement field may exhibit mixed features, including propagating behavior (Figs.~\ref{fig08}a,c), partial localization (Figs.~\ref{fig08}b,d), and quasi-cloaking-type behavior (Figs.~\ref{fig08}e,f), depending on the displacement component and on the type of incident wave.

Figure~\ref{fig08} illustrates this mixed response for a circular configuration with $N_p=12$. The selected value of $\Omega$ corresponds to a smooth local minimum of $\Gamma$ which satisfies the local-minimum condition but not the higher-order curvature criteria associated with trapping-dominated resonances. Accordingly, the resulting field does not display a fully localized trapping mode. Instead, different displacement components exhibit different qualitative behaviors, reflecting the transitional character of the case $H=1$ between the anomalous-dispersion regime, where trapping dominates, and the normal-dispersion regime, where cloaking-type responses become more prominent. This type of mixed response is observed not only at $H=1$, but also for values of $H$ approaching unity, and it consistently occurs for all pin numbers $N_{p}$ and Poisson’s ratios $\nu$ considered in this study.

\subsubsection{$H > 1$: Cloaking-type response}
\noindent
In the normal dispersion regime, $H>1$, the sharp trapping resonances observed for $H<1$ are strongly attenuated and the response becomes dominated by a cloaking-type behavior. In this regime, the pins weakly perturb the incident field, so that the displacement pattern resembles that of the incoming wave over the frequency range considered.

\begin{figure}[!htb]
\centering
\includegraphics[scale=0.45]{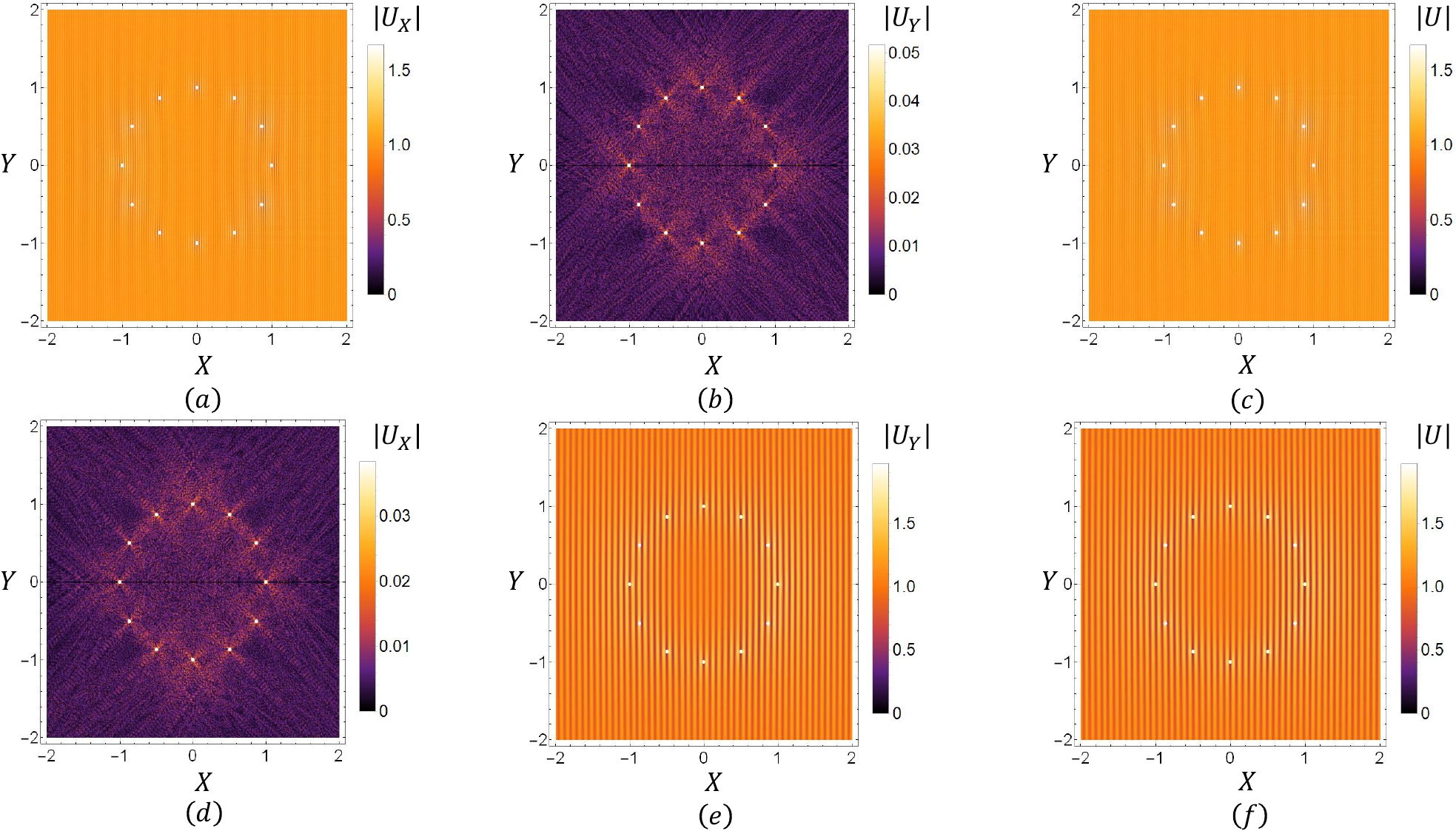}
\caption{Distribution of the non-dimensional displacement components and the magnitude of the non-dimensional displacement vector for a circular configuration, subjected to an incident $(a)-(c)$ P-wave and $(d)-(f)$ SV-wave of unit amplitude. In all cases, $H=10$, $\nu=0.25$, $\psi=0^{\circ}$, $\Omega=100$, $N_p = 12$.}
\label{fig09}
\end{figure}

As illustrated in Fig.~\ref{fig09}, the displacement component parallel to the incident polarization remains essentially unaffected by the presence of the pins, whereas the perpendicular component contains the scattered contribution. However, the latter remains several orders of magnitude smaller than the dominant component. The resulting displacement field is therefore almost one-dimensional, following primarily the polarization direction of the incident wave. Within the parameter ranges examined, this behavior is found to be weakly affected by the number of pins $N_p$, the Poisson's ratio $\nu$, and the non-dimensional frequency $\Omega$. Instead, the response is governed mainly by the pin geometry and by the normal-dispersion character of the medium. Thus, for $H>1$, the system does not exhibit pronounced trapping resonances; rather, the pins become effectively weak scatterers, giving rise to a cloaking-type response.

In the present study, the term cloaking is used in a qualitative sense to describe a response in which the presence of the pin configuration produces only a weak perturbation of the incident field. This notion is employed at a conceptual level and should be distinguished from transformation-based elastodynamic cloaking. It has been rigorously shown that perfect elastodynamic transformation cloaking is not achievable in strain gradient elastic solids  \cite{yavari2019nonlinear, sozio2021elastodynamic, sozio2023optimal}. Although coordinate transformations are not considered here, the weak-scattering response observed for $H>1$ exhibits a clear physical analogy with cloaking, in the sense that the pins become only weakly detectable through the displacement field.

\subsection{The effect of Poisson's ratio}
\label{sec5.3}
\noindent
The influence of Poisson's ratio on the system response is investigated using circular pin configurations with fixed $N_p$ and fixed microstructural ratio $H$. The analysis focuses first on the behavior of the logarithmic determinant $\Gamma$, since $\nu$ enters the Green's matrix through the ratio between the P and S-wave speeds $\beta$. As $\nu$ approaches the incompressible limit, $\nu\to 0.5$, the contrast between the P and S-wave speeds increases significantly, modifying the structure of the Green's matrix and, consequently, the distribution of local minima of $\Gamma$.

\begin{figure}[!htb]
\centering
\includegraphics[scale=0.7]{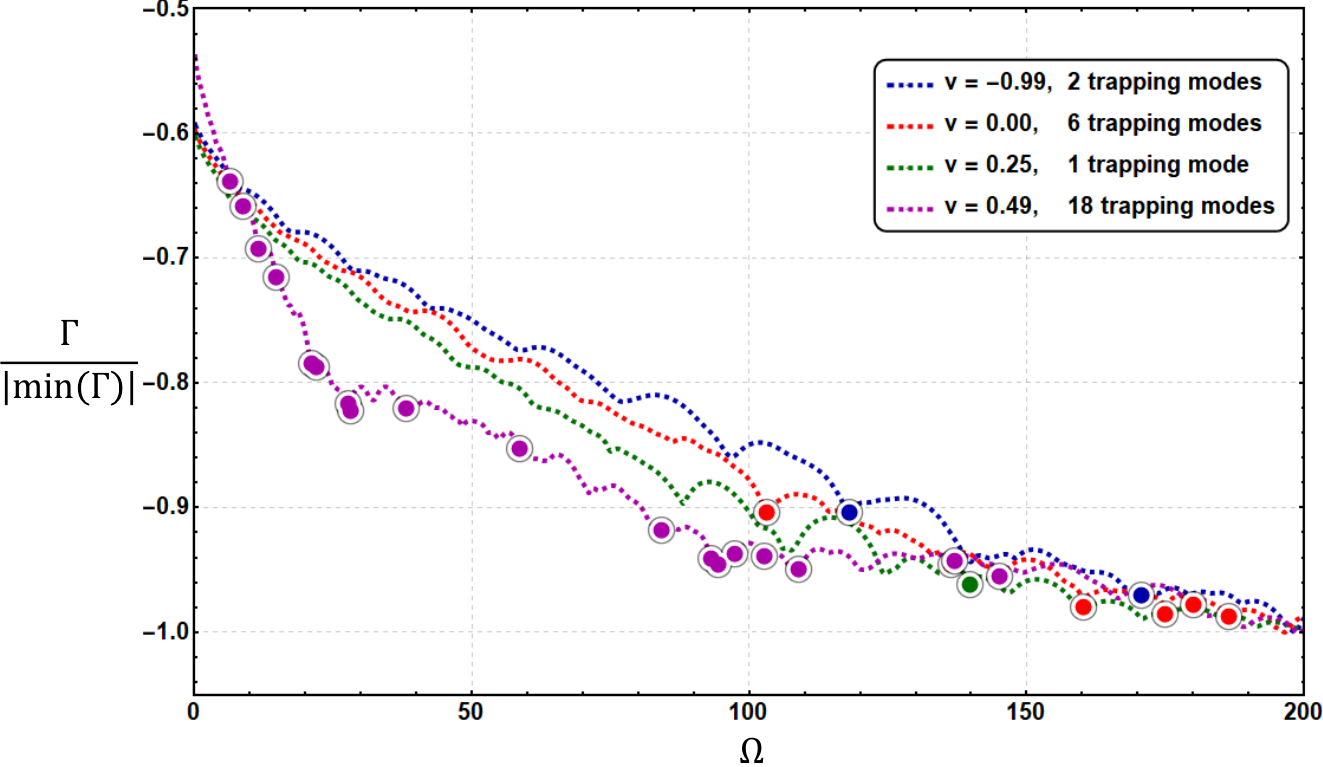}
\caption{Variation of the normalized determinant of the Green's matrix $\Gamma/|\min(\Gamma)|$ for a circular configuration, for representative values of Poisson's ratio, $H=0$, $N_p = 24$.}
\label{fig10}
\end{figure}

Figure~\ref{fig10} shows the variation of the normalized logarithmic determinant for representative values of $\nu$. For most values of Poisson's ratio, the number of local minima is only weakly affected. A notable exception occurs as $\nu$ approaches $0.5$, where a significant increase in the number of local minima is observed. This behavior persists across the circular configurations examined and indicates that near-incompressible materials exhibit increased sensitivity to resonant scattering phenomena.

Typical examples of materials with Poisson's ratio close to $\nu\simeq 0.5$ include rubber-like solids, elastomers, and biological soft tissues, whose bulk modulus is much larger than their shear modulus \cite{greaves2011poisson}. Architected materials, such as pentamode structures, may also approach this limit by design \cite{buckmann2012feasibility}. In contrast, auxetic materials $(\nu<0)$, including re-entrant foams, rotating-unit lattices, and certain mechanical metamaterials, exhibit a reduced P/SV wave-speed contrast and therefore provide a useful comparison for assessing the role of compressibility in the scattering response  \cite{lakes1987foam, milton1992composite, evans1991molecular, milton2013complete}.

\begin{figure}[!htb]
\centering
\includegraphics[scale=0.60]{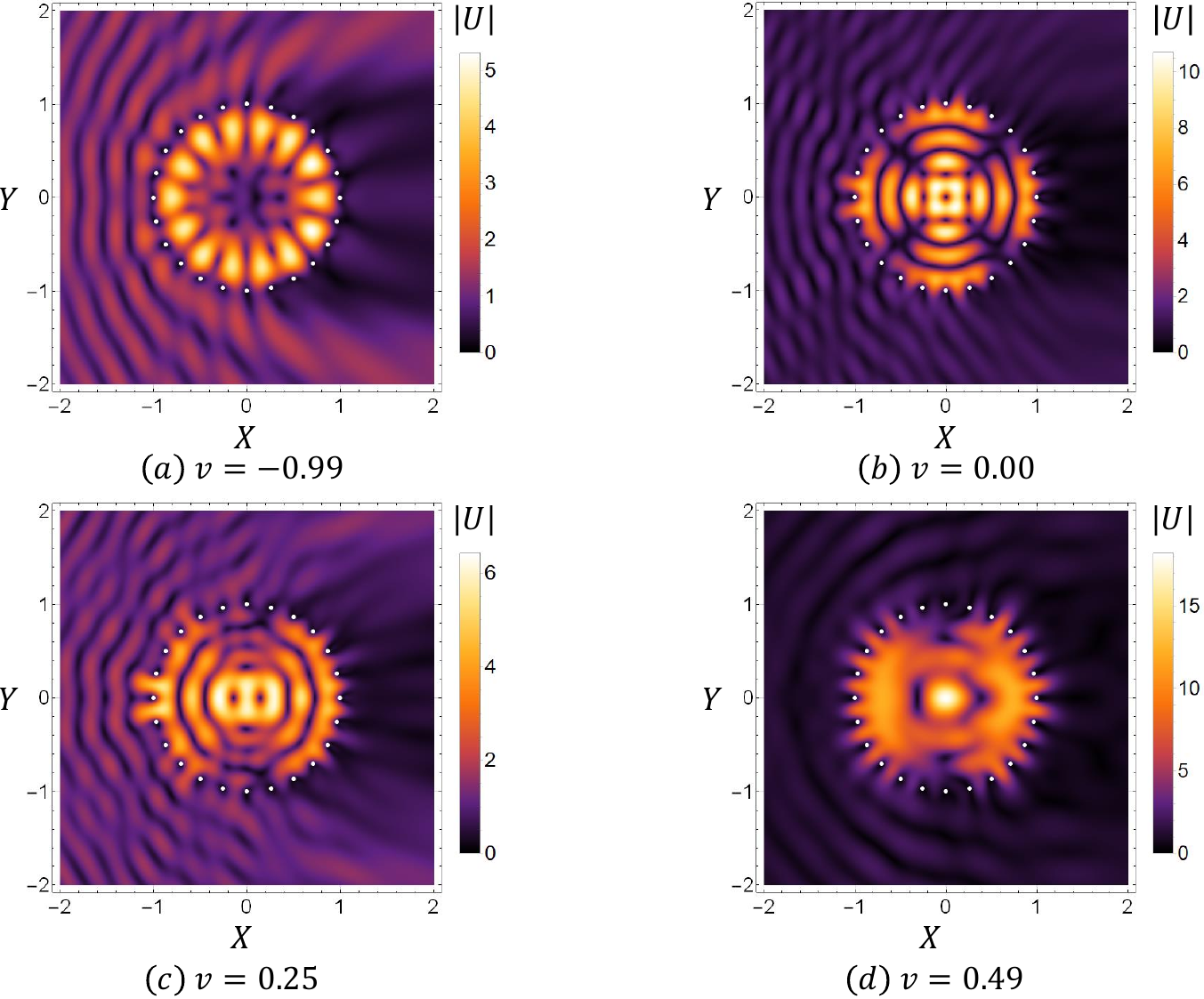}
\caption{Distribution of the magnitude of the non-dimensional displacement vector for a circular configuration, shown for different Poisson’s ratio values: $(a)$ $\Omega=118.03322$, SV incident wave; $(b)$ $\Omega=175.03035$, P incident wave; $(c)$ $\Omega=139.83396$, P incident wave; $(d)$ $\Omega=28.11431$, P incident wave. In all cases, $H=0$ and $\psi=0^{\circ}$, $N_p = 24$.}
\label{fig11}
\end{figure}

With respect to the displacement field, the computations show that the largest displacement amplitudes occur for $\nu=0.49$, regardless of the number of pins $N_p$, the value of $H$, or the type of incident plane wave considered. This behavior is illustrated in Fig.~\ref{fig11}, which presents the magnitude of the non-dimensional displacement vector for a circular array of $24$ pins at the trapping mode producing the strongest localization in the frequency range $0<\Omega<200$.

The results further indicate that Poisson's ratio influences which type of incident wave is more effective in exciting localized motion. For the cases examined here, the strongest localization in conventional materials $(\nu>0)$ is predominantly obtained under P-wave incidence, whereas in auxetic materials $(\nu<0)$ SV-wave incidence is more effective. This behavior should not be attributed solely to the relative values of the P and S-wave speeds. Rather, Poisson's ratio enters the Green's matrix through the parameter $\beta$ and modifies the coupling between the prescribed incident-field vector and the resonant response patterns of the pinned configuration. Thus, the observed dependence on the incident-wave type reflects the combined influence of the P/SV wave-speed contrast, the polarization of the incident field, and the structure of the localized modes.

It should also be noted that the incident waves are normalized here by displacement amplitude. A comparison based on equal incident energy flux could lead to a different assessment of the relative efficiency of P and SV-wave excitation, especially near the incompressible limit.
   

\subsection{The effect of wave incident angle}
\label{sec5.4}
\noindent
We now investigate the influence of the incidence angle $\psi$ of the incoming plane wave. For circular pin arrangements, the response is only weakly affected by $\psi$, especially when the number of pins is large. This behavior follows from the rotational symmetry of the configuration: in the continuous circular limit, changing the incidence angle simply rotates the displacement field. For a finite number of pins, this invariance is exact only for rotations compatible with the discrete symmetry of the array, but it remains a good approximation for sufficiently dense circular configurations.

The situation is different for pin arrangements lacking radial symmetry. In such cases, the incidence angle plays an important role in shaping both the scattering pattern and the extent of wave localization. Nevertheless, the determinant of the Green's tensor remains independent of $\psi$, since $\mathbf{G}$ depends only on the frequency, the material parameters, and the spatial arrangement of the pins. The incidence angle affects only the sampled incident-field vector $\tilde{\mathbf{U}}^{in}$, and therefore changes the way in which the incident wave excites the resonant response of the system. Consequently, the determinant-based candidate resonant frequencies remain unchanged, whereas the response amplitude, quantified by $\left\lvert U_{\max}\right\rvert$, varies with $\psi$.

\begin{figure}[!htb]
\centering
\includegraphics[scale=0.45]{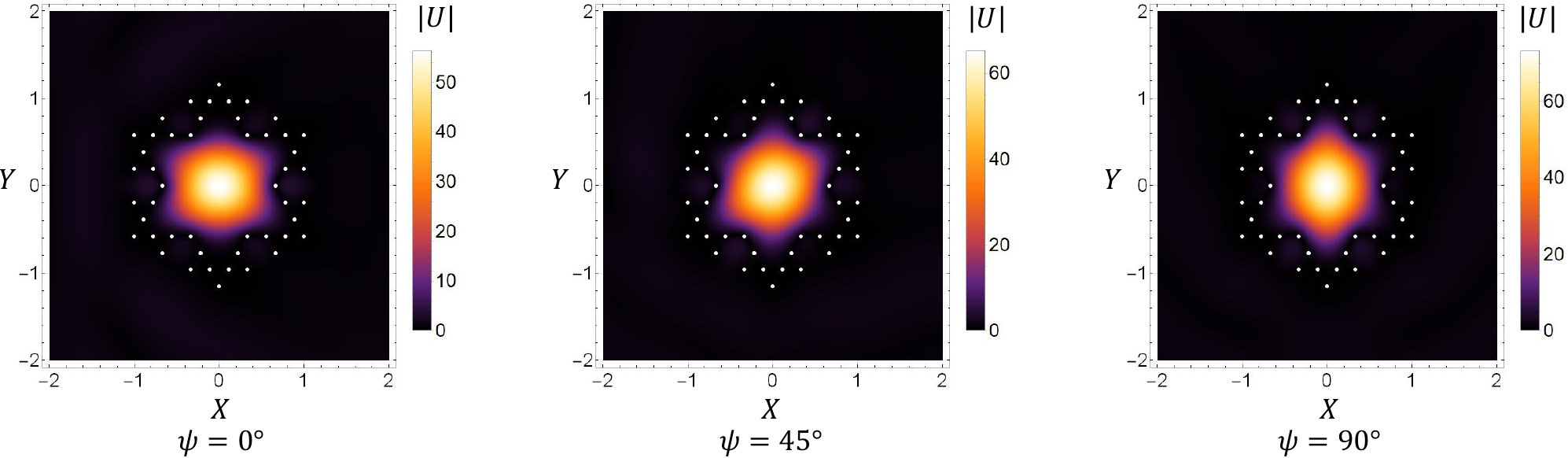}
\caption{Distribution of the magnitude of the non-dimensional displacement vector for an equilateral non-convex polygon, subjected to an incident P-wave, shown for different incidence angle values. In all cases, $H=0$, $\nu=0.25$, $\Omega=13.520056$, $N_p = 48$.}
\label{fig12}
\end{figure}

For each candidate resonance, an incidence angle can be identified that maximizes the response amplitude. In the non-radially symmetric configuration shown in Fig.~\ref{fig12}, the largest displacements occur when the incident plane wave impinges directly on a pin located at a convex corner relative to its neighbouring pins. For the equilateral non-convex polygon with $N_p=48$ considered here, the maximum displacement is obtained for $\psi=90^\circ$.

\subsection{The effect of multiple pin configurations}
\label{sec5.5}
\noindent
We finally investigate the response of compound pin arrangements formed by placing multiple circular configurations in close proximity. Two cases are considered: displaced circular configurations and offset circular configurations. These examples illustrate how interactions between neighbouring pinned regions can modify the trapping response and redistribute the localized displacement field.

\subsubsection{Multiple displaced circular pin configurations}
\noindent
For a single circular pin configuration, localized resonant motion is primarily confined within the region enclosed by the array. When several circular configurations are placed close to one another, additional interaction mechanisms arise. Depending on the resonant frequency, localization may occur simultaneously within several circular clusters, or it may be concentrated within only selected clusters, while the remaining regions become isolated.

\begin{figure}[!htb]
\centering
\includegraphics[scale=0.60]{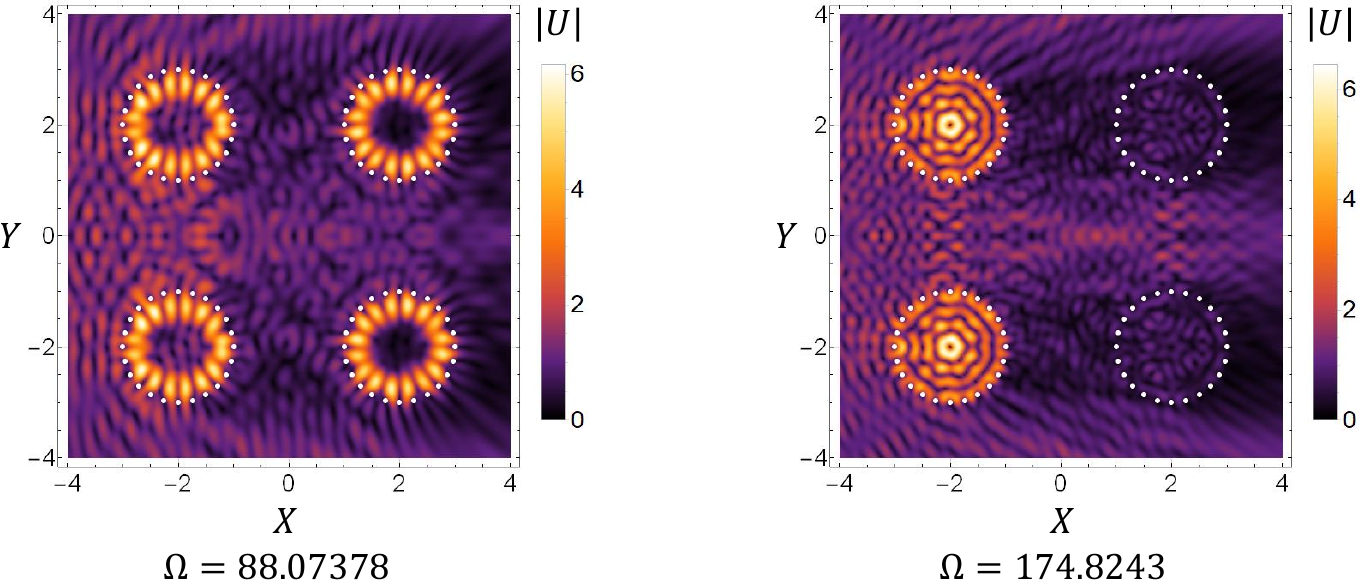}
\caption{Distribution of the magnitude of the non-dimensional displacement vector for multiple circular configurations subjected to an incident P-wave at different resonant frequency values. In all cases, $H=0$, $\nu=0.25$, $\psi=0^{\circ}$.}
\label{fig13}
\end{figure}

This behavior is illustrated in Fig.~\ref{fig13}. At the lower resonant frequency shown in Fig.~\ref{fig13}a, the displacement field localizes within all circular clusters. This response reflects the collective interaction of the incident and scattered waves with the compound configuration. At a higher resonant frequency shown in Fig.~\ref{fig13}b, localization is confined mainly to selected clusters, whereas other enclosed regions remain weakly excited. Thus, the compound arrangement can support both collective and selective trapping responses.

The distinction can be interpreted in terms of the wavelength relative to the inter-pin spacing and the separation between neighbouring clusters. At lower frequencies, the wavelength is large compared with the spacing between adjacent pins, and each circular array interacts with the wave field in a more collective manner. The scattered fields generated by the pins overlap strongly, promoting localization within several clusters. At higher frequencies, the wavelength becomes comparable to the characteristic spacing of the pins, and the discrete nature of the scatterers becomes more important. In this regime, multiple-scattering interference controls the localization pattern: constructive interference can enhance trapping within certain clusters, while destructive interference can suppress excitation in others.

\subsubsection{Circular pin arrangements with multiple offsets}
\begin{figure}[!htb]
\centering
\includegraphics[scale=0.45]{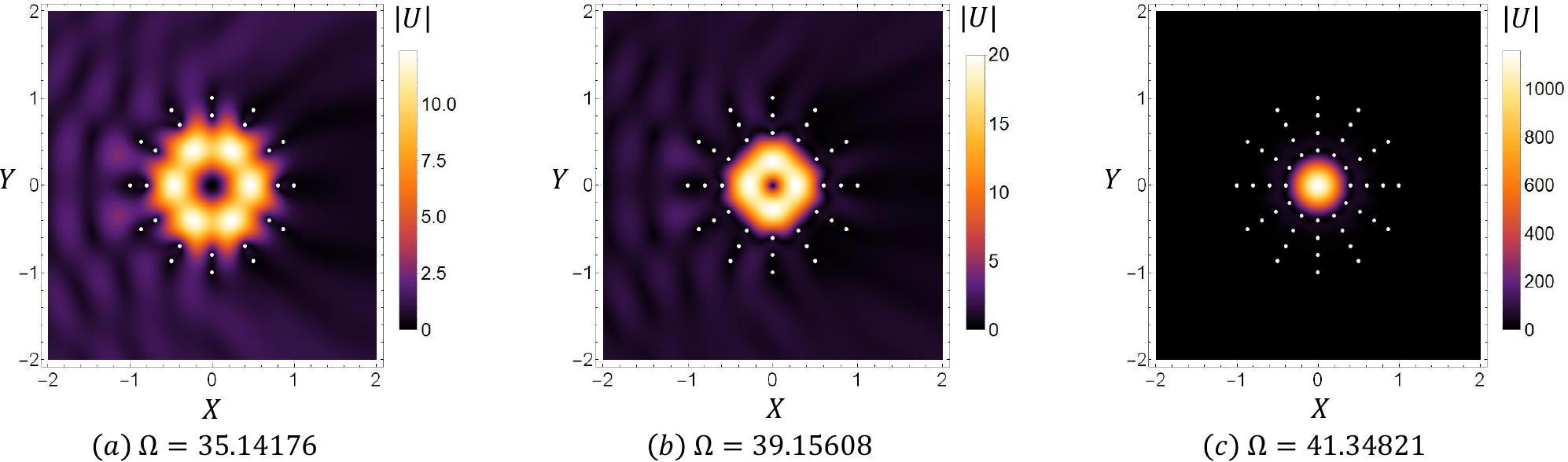}
\caption{Distribution of the magnitude of the non-dimensional displacement vector for multiple circular configurations subjected to an incident P-wave at close resonant frequencies: (a) 2 offsets, (b) 3 offsets, (c) 4 offsets, each with 12 pins. In all cases, $H=0$, $\nu=0.25$, $\psi=0^{\circ}$, and offset spacing $=0.2$.}
\label{fig14}
\end{figure}

\noindent
Among the pin configurations examined, the strongest displacement localization is observed when multiple offsets are introduced. As shown in Fig.~\ref{fig14}, increasing the number of offset circular arrays leads to a progressive increase in the maximum displacement amplitude at nearby resonant frequencies. This indicates that offset configurations can enhance trapping by promoting repeated scattering and constructive interference within the pinned region.

This amplification may be interpreted as a consequence of the modified propagation paths created by the successive offsets. Each offset alters the spatial distribution of the scattered field, so that repeated interactions between neighbouring circular arrays can reinforce the displacement within the compound configuration. As the number of offsets increases, the confinement of the displacement field becomes more pronounced, leading to stronger localization near the selected resonant frequencies.

\begin{figure}[!htb]
\centering
\includegraphics[scale=0.45]{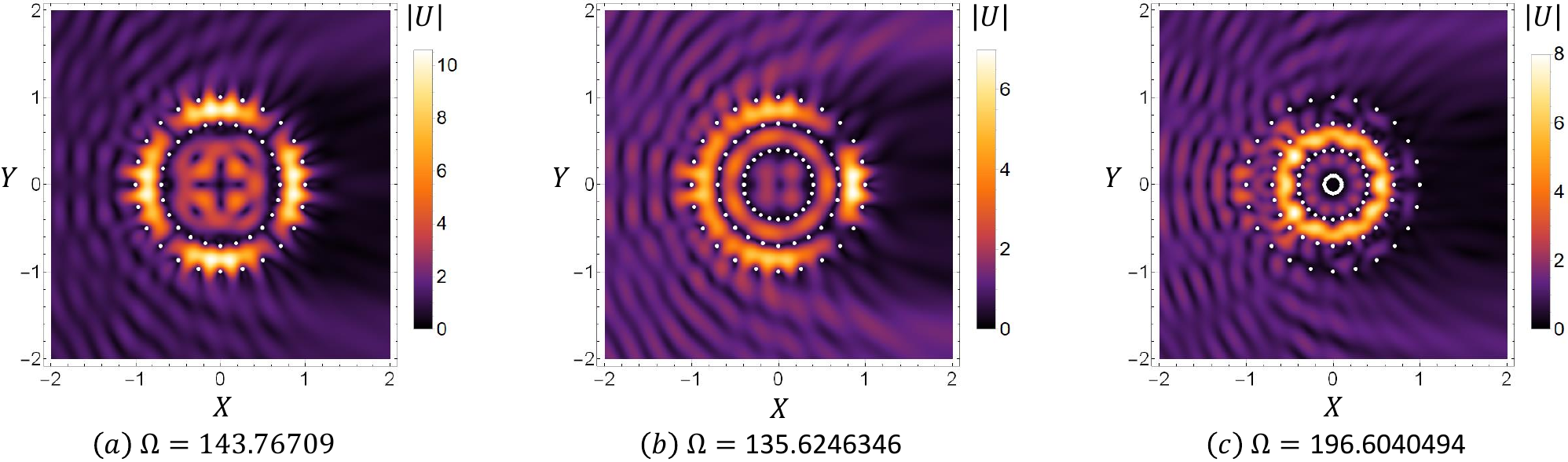}
\caption{Distribution of the magnitude of the non-dimensional displacement vector for multiple circular configurations subjected to an incident P-wave at resonant frequencies: (a) 2 offsets, (b) 3 offsets, (c) 4 offsets, each with 24 pins. In all cases, $H=0$, $\nu=0.25$, $\psi=0^{\circ}$, and offset spacing $=0.3$.}
\label{fig15}
\end{figure}

Finally, adjusting the offset spacing reveals that at specific resonant frequencies, maximum trapping occurs between the outer offsets rather than within the inner circle, as shown in Fig.~\ref{fig15}. For the circular configurations considered here, this behavior is observed when the normalized offset spacing exceeds $0.3$.

\section{Concluding remarks}
\label{sec6}		
\noindent 
This study investigated the scattering of plane waves by zero-area rigid point constraints (pins) embedded in an infinite plane-strain medium governed by strain gradient elasticity. A central contribution is the derivation of a closed-form time-harmonic Green's tensor for the plane-strain problem. Owing to the regularizing character of the strain gradient theory, the displacement Green's tensor remains bounded at the source, which allows point scatterers to be treated directly through a superposition of fundamental solutions. This provides the basis for extending point-scatterer methods, originally developed for scalar Kirchhoff-plate problems, to the fully vectorial plane-strain setting.

The scattering problem was formulated as a non-homogeneous algebraic system for the reaction amplitudes associated with the pins. Resonant responses were identified through the frequency-dependent behavior of the Green's matrix. In particular, the determinant of the Green's tensor was used to locate candidate resonant frequencies through local minima, while higher-order curvature criteria were employed to distinguish trapping-dominated resonances from ordinary non-localized scattering responses. This distinction is important because not every local minimum of $\Gamma$ leads to a localized mode.

The results show that the ratio between the microinertial length $h$ and the energetic strain-gradient length $\ell$ plays a decisive role in the response. In the anomalous dispersion regime, $H=h/\ell<1$, sharp resonant minima are observed and the displacement field may become strongly localized within the pin configuration. In contrast, in the normal dispersion regime, $H>1$, these trapping resonances are strongly attenuated and the response becomes dominated by a cloaking-type, weak-scattering behavior, in which the pins only weakly perturb the incident wave field over the frequency range considered. The transitional case $H=1$ combines non-dispersive propagating waves with the regularizing contribution of the strain gradient Green's tensor, leading to mixed responses involving propagation, partial localization, and quasi-cloaking behavior.

The geometry of the pin configuration was also found to play a central role. Circular arrays support localized trapping modes, with denser configurations and higher frequencies promoting stronger multiple-scattering interactions. In some cases, especially for dense arrays, the trapped field becomes concentrated along the pinned perimeter, indicating the formation of a perimeter-localized trapping mode resembling a whispering gallery. Compound arrangements involving multiple displaced or offset circular configurations further enrich the response, producing collective or selective localization depending on the resonant frequency and the spatial arrangement of the clusters. Offset configurations, in particular, can act as effective waveguiding structures that enhance confinement within the pinned region.

Poisson's ratio affects both the structure of the Green's matrix and the coupling between the incident field and the localized response. The computations indicate that, as $\nu$ approaches the incompressible limit, the number and strength of resonant responses increase. The relative effectiveness of P and SV-wave incidence also depends on $\nu$, reflecting the combined influence of wave-speed contrast, polarization, and the spatial structure of the resonant modes.

Overall, the analysis demonstrates that strain gradient elasticity provides a mechanism for tuning scattering, trapping, and cloaking-type responses through the interplay between material length scales and pin geometry. The results suggest possible routes for designing microstructured media and elastic metamaterials with tailored wave-localization and wave-screening properties.

\begin{appendices}
\numberwithin{figure}{section}
\numberwithin{equation}{section}
\section[Derivation of the Green's function]{Derivation of the Green's function}
\label{A}
\noindent
To solve the above system of partial differential equations \eqref{eq:3.7} and \eqref{eq:3.8}, the double Fourier transform is employed, defined as follows
\begin{equation}\label{eq:A.1}
\bar{f}(\xi,\eta) = \int_{-\infty}^{+\infty} \int_{-\infty}^{+\infty} f(x,y) e^{i (\xi x + \eta y)} \,dx\,dy.
\end{equation}
The inverse double Fourier transform is respectively defined as
\begin{equation}\label{eq:A.2}
f(x,y) = \frac{1}{4 \pi^2} \int_{-\infty}^{+\infty} \int_{-\infty}^{+\infty} \bar{f}(\xi,\eta) e^{-i (\xi x + \eta y)} \,d\xi\,d\eta.
\end{equation}
For simplicity, and without loss of generality, we assume that the concentrated body force $F_j$ is applied at the origin i.e., $F_i = p_i \delta(x) \delta(y)$, $j = x, y$. A coordinate system shift will then be performed, allowing the Green's function to be extended to the general case where the body force is applied at any arbitrary point.

Applying the double Fourier transform to Eqs.~\eqref{eq:3.7} and \eqref{eq:3.8} yields the following system of algebraic equations in the transform space
\begin{equation}\label{eq:A.3}
\begin{bmatrix} 
A\xi^2 + C\eta^2 - D & B\xi\eta \\ 
B\xi\eta & C\xi^2 + A\eta^2 - D 
\end{bmatrix} 
\begin{bmatrix} 
\bar{u}_x \\ 
\bar{u}_y 
\end{bmatrix} 
=
\begin{bmatrix} 
p_x \\ 
p_y 
\end{bmatrix},
\end{equation}
where
\begin{equation}\label{eq:A.4}
A = (\lambda + 2\mu) a_{\ell}^2, \quad 
B = (\lambda + \mu) a^2_{\ell}, \quad
C = \mu a^2_{\ell}, \quad 
D = \rho \omega^2 a^2_{h},
\end{equation}
where the functions $a_{\ell}$ and $a_{h}$ are defined by
\begin{equation}\label{eq:A.5}
a_{\ell}^2 \equiv a_{\ell}^2(q) = 1 + \ell^2 q^2, \quad
a_{h}^2 \equiv a_{h}^2(q) = 1 + h^2 q^2, \quad
q^2 = \xi^2 + \eta^2
\end{equation}
From Eqs.~\eqref{eq:A.3}-\eqref{eq:A.5} the following expressions for the transformed displacement components are obtained:
\begin{equation}\label{eq:A.6}
\bar{u}_x = \frac{\left[ \left( \xi^2 + \beta^2 \eta^2 \right) a_{\ell}^2 - a_{h}^2 \beta^2 k^2_P \right] p_x
- \left( \beta^2 -1 \right) \xi \eta a_{\ell}^2 p_y}{\mu \beta^2 D_P D_S},
\end{equation}
\begin{equation}\label{eq:A.7}
\bar{u}_y = \frac{\left[ \left( \beta^2 \xi^2 + \eta^2 \right) a_{\ell}^2 - a_{h}^2 \beta^2 k^2_P \right] p_y
- \left( \beta^2 -1 \right) \xi \eta a_{\ell}^2 p_x}{\mu \beta^2 D_P D_S}.
\end{equation}
where the quantity $\beta$ is defined in Eq.~\eqref{eq:3.14}. The quantities $D_j$ are defined as
\begin{equation}\label{eq:A.8}
D_j \equiv D_j(\omega, q)  = a_{\ell}^2 q^2 - a_{h}^2 k_j^2, \quad k_j = \frac{\omega}{c_j}, \quad j = P, S.
\end{equation}
Setting $D_j=0$ provides the dispersion relations for the pressure and shear waves given by Eqs.~\eqref{eq:2.14} and \eqref{eq:2.15}.

We proceed by presenting the inversion of $\bar{u}_x$ and $\bar{u}_y$. The starting point is decomposing the transformed expressions as follows:
\begin{equation}\label{eq:A.9}
\bar{u}_x = p_x \, \bar{g}_{xx} + p_y \, \bar{g}_{xy},
\end{equation}
\begin{equation}\label{eq:A.10}
\bar{u}_y = p_x \, \bar{g}_{yx} + p_y \, \bar{g}_{yy},
\end{equation}
where $\bar{g}_{ij}$ denotes the transformed displacement in the $i-$direction due to a force of unit magnitude in the $j-$direction. After some extensive but straightforward algebra the following expressions are obtained
\begin{equation}\label{eq:A.11}
\begin{gathered}
\bar{g}_{xx}
= \frac{1}{\mu \beta^2 q^2}
\left( \frac{\xi^2}{D_P} + \frac{\beta^2 \eta^2}{D_S} \right),
\\[6pt]
\bar{g}_{xy}=\bar{g}_{yx}
= \frac{\xi \eta}{\mu \beta^2 q^2}
\left( \frac{1}{D_P} - \frac{\beta^2}{D_S} \right),
\qquad
\bar{g}_{yy}
= \frac{1}{\mu \beta^2 q^2}
\left( \frac{\eta^2}{D_P}+\frac{\beta^2 \xi^2}{D_S}  \right).
\end{gathered}
\end{equation}
Applying the inverse double Fourier transform \eqref{eq:A.2} to Eqs.~\eqref{eq:A.11} yields the components of the Green's tensor
\begin{equation}\label{eq:A.13}
g_{xx} = \frac{1}{4 \pi^2 \mu \beta^2} \Bigg[ 
\int_{-\infty}^{+\infty} \int_{-\infty}^{+\infty} 
\frac{\xi^2 e^{-i (\xi x + \eta y)}}{q^2 D_P} \,d\xi\,d\eta
+ \int_{-\infty}^{+\infty} \int_{-\infty}^{+\infty} 
\frac{\beta^2 \eta^2 e^{-i (\xi x + \eta y)}}{q^2 D_S} \,d\xi\,d\eta \Bigg],
\end{equation}
\begin{equation}\label{eq:A.14}
g_{xy} =g_{yx}= \frac{1}{4 \pi^2 \mu \beta^2} \Bigg[ 
\int_{-\infty}^{+\infty} \int_{-\infty}^{+\infty} 
\frac{\xi \eta e^{-i (\xi x + \eta y)}}{q^2 D_P} \,d\xi\,d\eta 
- \int_{-\infty}^{+\infty} \int_{-\infty}^{+\infty} 
\frac{\beta^2 \xi \eta e^{-i (\xi x + \eta y)}}{q^2 D_S} \,d\xi\,d\eta \Bigg],
\end{equation}
\begin{equation}\label{eq:A.15}
g_{yy}= \frac{1}{4 \pi^2 \mu \beta^2} \Bigg[ 
\int_{-\infty}^{+\infty} \int_{-\infty}^{+\infty} 
\frac{\beta^2 \xi^2 e^{-i (\xi x + \eta y)}}{q^2 D_S} \,d\xi\,d\eta
+ \int_{-\infty}^{+\infty} \int_{-\infty}^{+\infty} 
\frac{\eta^2 e^{-i (\xi x + \eta y)}}{q^2 D_P} \,d\xi\,d\eta \Bigg].
\end{equation}
To facilitate the evaluation of the integrals in Eqs.~\eqref{eq:A.13}-\eqref{eq:A.15}, it is convenient to convert the spatial coordinates $(x,y)$ and the wavenumber components $(\xi,\eta)$ into polar coordinates. The transformations are given by:
\begin{equation}\label{eq:A.16}
x = r \cos(\theta), \quad y = r \sin(\theta), \quad r^2 = x^2 + y^2,
\end{equation}
\begin{equation}\label{eq:A.17}
\xi = q \cos(\phi), \quad \eta = q \sin(\phi), \quad q^2 = \xi^2 + \eta^2.
\end{equation}
Using Eqs.~\eqref{eq:3.14}, \eqref{eq:A.16}, \eqref{eq:A.17}, along with the Jacobian of the transformation $J = q$, Eqs.~\eqref{eq:A.13}-\eqref{eq:A.15} can be rewritten as
\begin{align}\label{eq:A.18}
g_{xx} &= \frac{1}{4 \pi^2 \mu \beta^2} \Bigg[ 
\int_{0}^{\infty} \int_{0}^{ 2 \pi} 
\frac{q \cos^2(\phi)}{D_P} e^{-i q r [\cos(\theta - \phi)]} \,d\phi\,dq \nonumber \\
&+ \beta^2 \int_{0}^{\infty} \int_{0}^{2 \pi} 
\frac{q \sin^2(\phi)}{D_S} e^{-i q r [\cos(\theta - \phi)]} \,d\phi\,dq \Bigg],
\end{align}
\begin{align}\label{eq:A.19}
g_{xy} &= \frac{1}{4 \pi^2 \mu \beta^2} \Bigg[ 
\int_{0}^{\infty} \int_{0}^{2 \pi} 
\frac{q \sin(\phi) \cos(\phi)}{D_P} e^{-i q r [\cos(\theta - \phi)]} \,d\phi\,dq \nonumber \\
&- \beta^2 \int_{0}^{\infty} \int_{0}^{2 \pi} 
\frac{q \sin(\phi) \cos(\phi)}{D_S} e^{-i q r [\cos(\theta - \phi)]} \,d\phi\,dq \Bigg],
\end{align}
\begin{align}\label{eq:A.20}
g_{yy} &= \frac{1}{4 \pi^2 \mu \beta^2} \Bigg[ 
\beta^2 \int_{0}^{\infty} \int_{0}^{ 2 \pi} 
\frac{q \cos^2(\phi)}{D_S} e^{-i q r [\cos(\theta - \phi)]} \,d\phi\,dq \nonumber \\
&+ \int_{0}^{\infty} \int_{0}^{2 \pi} 
\frac{q \sin^2(\phi)}{D_P} e^{-i q r [\cos(\theta - \phi)]} \,d\phi\,dq \Bigg],
\end{align}
where for the derivation of Eqs.~\eqref{eq:A.18}-\eqref{eq:A.20} the following well-known trigonometric identity has been utilized
\begin{equation}\label{eq:A.21}
\cos(\phi) \cos(\theta) + \sin(\phi) \sin(\theta) = \cos(\phi - \theta).
\end{equation}
Eqs.~\eqref{eq:A.18}-\eqref{eq:A.20} can be further simplified using the following well-known identities
\begin{equation}\label{eq:A.22}
\int_{0}^{2 \pi} \sin^2(\phi) e^{-i q r [\cos(\theta - \phi)]} \,d\phi\ =
2 \pi \left[ J_0(q r) \sin^2(\theta) + \frac{J_1(q r)}{q r} \cos(2 \theta) \right],
\end{equation}
\begin{equation}\label{eq:A.23}
\int_{0}^{2 \pi} \cos^2(\phi) e^{-i q r [\cos(\theta - \phi)]} \,d\phi\ =
2 \pi \left[ J_0(q r) \cos^2(\theta) - \frac{J_1(q r)}{q r} \cos(2 \theta) \right],
\end{equation}
\begin{equation}\label{eq:A.24}
\int_{0}^{2 \pi} \sin(\phi) \cos(\phi) e^{-i q r [\cos(\theta - \phi)]} \,d\phi\ =
2 \pi \sin(\theta) \cos(\theta) \left[ J_0(q r) - \frac{2 J_1(q r)}{q r} \right],
\end{equation}
where $J_n(q r)$ is the Bessel function of the first kind of order $n$.

By substituting Eqs.~\eqref{eq:A.22}-\eqref{eq:A.24} into \eqref{eq:A.18}-\eqref{eq:A.20} we obtain
\begin{equation}\label{eq:A.25}
g_{xx} = \frac{1}{2 \pi \mu \beta^2} \left[ \cos^2(\theta) I^{(P)}_0 + \beta^2 \sin^2(\theta) I^{(S)}_0 - \frac{\cos(2 \theta)}{r} \left( I^{(P)}_1 - \beta^2 I^{(S)}_1 \right) \right],
\end{equation}
\begin{equation}\label{eq:A.26}
g_{xy} = \frac{1}{2 \pi \mu \beta^2} \sin(\theta) \cos(\theta) \left[ I^{(P)}_0 - \beta^2 I^{(S)}_0 - \frac{2}{r} \left( I^{(P)}_1 - \beta^2 I^{(S)}_1 \right) \right],
\end{equation}
\begin{equation}\label{eq:A.27}
g_{yy}= \frac{1}{2 \pi \mu \beta^2} \left[ \sin^2(\theta) I^{(P)}_0 + \beta^2 \cos^2(\theta) I^{(S)}_0 + \frac{\cos(2 \theta)}{r} \left( I^{(P)}_1 - \beta^2 I^{(S)}_1 \right) \right],
\end{equation}
where
\begin{equation}\label{eq:A.28}
I^{(j)}_0(r) = \int_{0}^{\infty} \frac{q}{D_j} J_0(q r) \,dq\,, \quad j = P, S,
\end{equation}
\begin{equation}\label{eq:A.29}
I^{(j)}_1(r) = \int_{0}^{\infty} \frac{1}{D_j} J_1(q r) \,dq\,, \quad j = P, S.
\end{equation}
The integrals in Eqs.~\eqref{eq:A.28}, \eqref{eq:A.29} can be evaluated using the following result
\begin{equation}\label{eq:A.30}
\frac{1}{D_j} = \frac{1}{\ell^2 \left( q^2_{1j} + q^2_{2j} \right)} \left( \frac{1}{q^2 - q^2_{1j}} - \frac{1}{q ^2 + q^2_{2j}} \right), \quad j = P, S,
\end{equation}
where the expressions of $q_{1j}$ and $q_{2j}$ are given by equations \eqref{eq:3.15} and \eqref{eq:3.16}.

Furthermore, with the outgoing-wave radiation prescription, the following identities hold \cite{graff2012wave}:
\begin{equation}\label{eq:A.31}
\int_{0}^{\infty} \frac{q}{q ^2 - q^2_{1j}} J_0(q r) \,dq\, = \frac{\pi i}{2} H^{(1)}_0(q_{1j} r), \quad j = P, S,
\end{equation}
\begin{equation}\label{eq:A.32}
\int_{0}^{\infty} \frac{q}{q ^2 + q^2_{2j}} J_0(q r) \,dq\, = K_0(q_{2j} r) , \quad j = P, S,
\end{equation}
\begin{equation}\label{eq:A.33}
\int_{0}^{\infty} \frac{1}{q ^2 - q^2_{1j}} J_1(q r) \,dq\, = \frac{\pi i}{2} \frac{H^{(1)}_1(q_{1j} r)}{q_{1j}} - \frac{1}{q^2_{1j} r}, \quad j = P, S,
\end{equation}
\begin{equation}\label{eq:A.34}
\int_{0}^{\infty} \frac{1}{q ^2 + q^2_{2j}} J_1(q r) \,dq\, = \frac{1}{q^2_{2j} r} - \frac{K_1(q_{2j} r)}{q_{2j}}, \quad j = P, S.
\end{equation}
With the help of Eqs.~\eqref{eq:A.31}-\eqref{eq:A.34}, Eqs.~\eqref{eq:A.28} and \eqref{eq:A.29} take the form
\begin{equation}\label{eq:A.35}
I^{(j)}_0 = \frac{1}{\ell^2 \left( q^2_{1j} + q^2_{2j} \right)} \left[ \frac{\pi i}{2} H^{(1)}_0(q_{1j} r) - K_0(q_{2j} r) \right], \quad j = P, S,
\end{equation}
\begin{equation}\label{eq:A.36}
I^{(j)}_1 = \frac{1}{\ell^2 \left( q^2_{1j} + q^2_{2j} \right)} \left[ - \left( \frac{1}{q^2_{1j}} + \frac{1}{q^2_{2j}} \right) \frac{1}{r} + \frac{\pi i}{2} \frac{H^{(1)}_1(q_{1j} r)}{q_{1j}} + \frac{K_1(q_{2j} r)}{q_{2j}} \right], \quad j = P, S.
\end{equation}
The expressions of the displacement Green's tensor components are given in Eqs.~\eqref{eq:3.9}-\eqref{eq:3.11} are finally obtained by substituting Eqs.~\eqref{eq:A.35} and \eqref{eq:A.36} into \eqref{eq:A.25}-\eqref{eq:A.27} and then applying the following identities \cite{watson1922treatise, abramowitz1965handbook}
\begin{equation}\label{eq:A.37}
\frac{2}{r} \frac{H^{(1)}_1(q_{1j} r)}{q_{1j}}- H^{(1)}_0(q_{1j} r) = H^{(1)}_2(q_{1j} r), \quad j = P, S,
\end{equation}
\begin{equation}\label{eq:A.38}
\frac{2}{r} \frac{K_1(q_{2j} r)}{q_{2j}}+K_0(q_{2j} r) = K_2(q_{2j} r), \, \, \quad j = P, S.
\end{equation}
It is noted that Eqs.~\eqref{eq:3.9}--\eqref{eq:3.11} are obtained from the expressions above by translating the source from the origin to the point $\mathbf{r}'=(x',y')$, so that $r$ is replaced by $s=|\mathbf{r}-\mathbf{r}'|$.

\begin{figure}[!htb]
\centering
\includegraphics[scale=0.45]{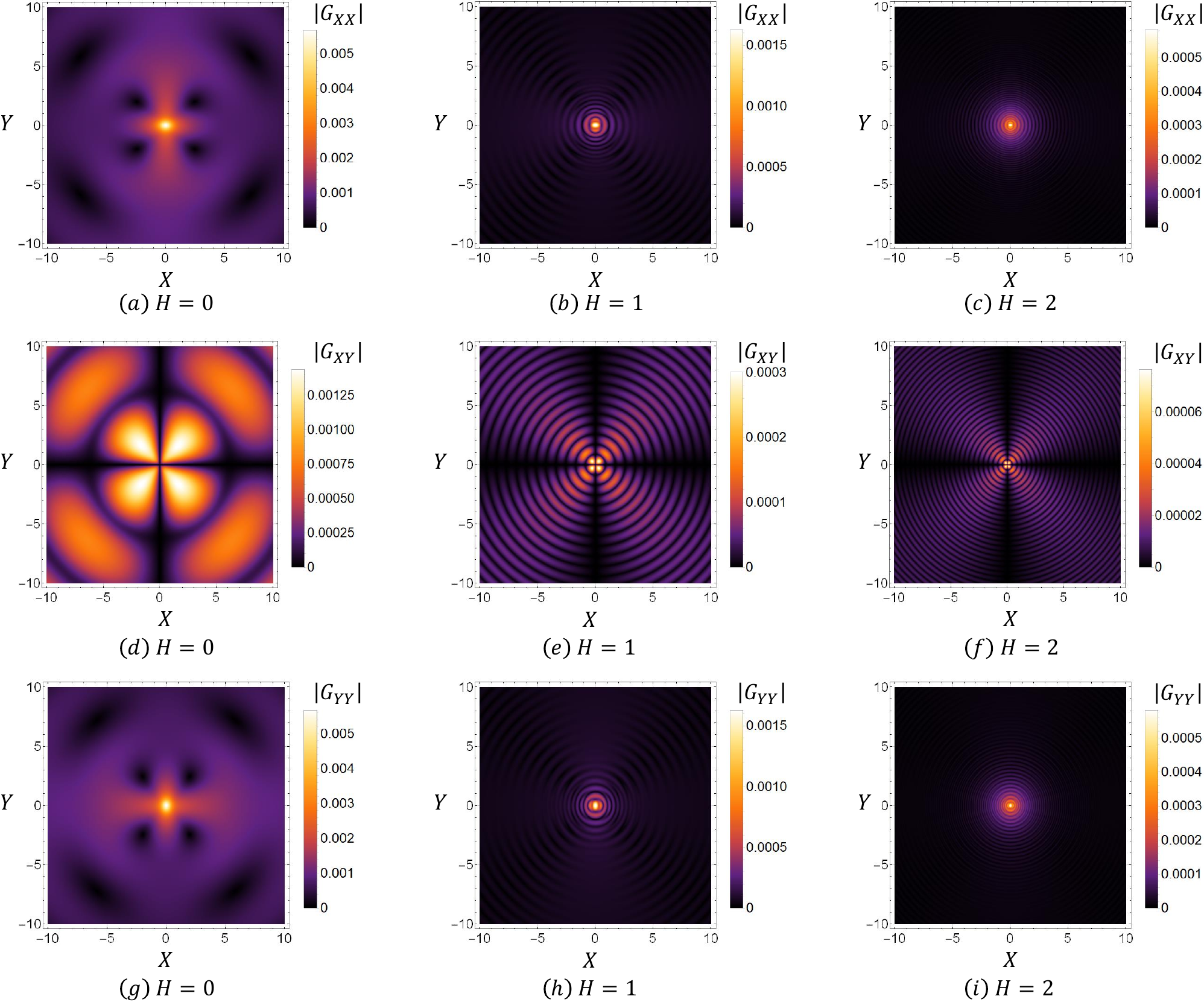}
\caption{Distribution of the non-dimensional Green's function components. In all cases, $\Omega=10$, $\nu=0.25$.}
\label{fig16}
\end{figure}

The distribution of the non-dimensional Green's function tensor is shown in Fig.~\ref{fig16}, using the normalization defined in Eq.~\eqref{eq:4.5}. In particular, it is observed that the Green's function remains bounded at the origin, consistent with its asymptotic behavior in Eq.~\eqref{eq:3.19}. As the parameter $H$ increases, displacement localization becomes less pronounced, consistent with the behavior observed in the scattering problems discussed earlier.

\end{appendices}

\section*{Acknowledgments}
\addcontentsline{toc}{section}{Acknowledgments}
E. Alevras was financially supported by \textgreek{ΕΛΚΕ} of the National Technical University of Athens.

\bibliographystyle{elsarticle-num}
\bibliography{References.bib}

\end{document}